\documentclass[pdftex,twocolumn,epjc3]{svjour3}          

\RequirePackage[T1]{fontenc}

\smartqed  

\RequirePackage{graphicx}
\usepackage{latexsym}
\usepackage{amssymb, amsmath, } 
\usepackage{subcaption}
\RequirePackage{mathptmx}      
\RequirePackage{flushend}
\RequirePackage[numbers,sort&compress]{natbib}
\RequirePackage[colorlinks,citecolor=blue,urlcolor=blue,linkcolor=blue]{hyperref}

\journalname{Eur. Phys. J. C}

\begin{document}

\title{Stellar equilibrium configurations of white dwarfs in the $f(R,T)$ gravity}


\author{G. A. Carvalho\thanksref{e1,addr1}
        \and
        R. V. Lobato\thanksref{addr1,addr2,addr3}
        \and
        P. H. R. S. Moraes\thanksref{addr1}
        \and
        Jos\'e D. V. Arba\~nil\thanksref{addr4}
        \and
        E. Otoniel\thanksref{addr5}
        \and
        R. M. Marinho Jr\thanksref{addr1}  
        \and
        M. Malheiro\thanksref{addr1}
}

\thankstext{e1}{e-mail: araujogc@ita.br}

\institute{Departamento de F\'isica, Instituto Tecnol\'ogico de Aeron\'autica, S\~ao Jos\'e dos Campos, SP, 12228-900, Brazil\label{addr1}
          \and
          Dipartimento di Fisica, Sapienza Universit\`a di Roma, P.le Aldo Moro 5, I-00185 Rome, Italy\label{addr2}
          \and
          ICRANet, P.zza della Repubblica 10, I-65122 Pescara, Italy\label{addr3}
          \and
          Departamento de Ciencias, Universidad Privada del Norte,
Avenida Alfredo Mendiola 6062 Urbanizaci\'on Los Olivos, Lima, Peru\label{addr4}
          \and
          Instituto de Forma\c{c}\~ao de Professores, Universidade Federal do Cariri, Brejo Santo, CE, 63260-000, Brazil \label{addr5}
}

\date{Received: date / Accepted: date}

\maketitle

\begin{abstract}
In this work we investigate the equilibrium configurations of white dwarfs in a modified gravity theory, na\-mely, $f(R,T)$ gravity, for which $R$ and $T$ stand for the Ricci scalar and trace of the energy-momentum tensor, respectively. Considering the functional form $f(R,T)=R+2\lambda T$, with $\lambda$ being a constant, we obtain the hydrostatic equilibrium equation for the theory. Some physical properties of white dwarfs, such as: mass, radius, pressure and energy density, as well as their dependence on the parameter $\lambda$ are derived. More massive and larger white dwarfs are found for negative values of $\lambda$ when it decreases. The equilibrium configurations predict a maximum mass limit for white dwarfs slightly above the Chandrasekhar limit, with larger radii and lower central densities when compared to standard gravity outcomes. The most important effect of $f(R,T)$ theory for massive white dwarfs is the increase of the radius in comparison with GR and also $f(R)$ results. By comparing our results with some observational data of massive white dwarfs we also find a lower limit for $\lambda$, namely, $\lambda >- 3\times 10^{-4}$. 
\end{abstract}

\section{Introduction}
\label{sec:intro}

White dwarfs (WDs) are the final evolution state of main sequence stars with initial masses up to $8.5-10.6M_{\odot}$. They correspond to 95-97\% of all observed stars in the Universe \cite{Woosley2015}. The main sequence progenitors can reach sufficiently high core temperatures ($8-12\times10^{8}$ K), to proceed to carbon burning and produce either oxygen-neon (ONe) core WDs or undergo a core-collapse supernova (SNII) via electron capture on the products of carbon burning. 

Chandrasekhar has shown that a WD cannot sustain a mass over $1.44M_{\odot}$, establishing the so-called Chandrasekhar mass limit \cite{Chandrasekhar1931}. If a WD grows over this limit, as in binary systems in which a WD is receiving mass from a nearby star, a type Ia supernova (SNIa) explosion may occur. SNIa progenitors are expected to be similar, with nearly equal luminosity, therefore being considered standard candles \cite{Weinberg2008}. In fact, in the late 1990s, the use of SNIa led to the discovery that the expansion of the Universe is accelerating \cite{Riess1998,Perlmutter1999}.

Nevertheless, some super luminous SNIas were found recently \cite{AndrewHowell2006, Scalzo2010}. It has been suggested that their progenitors are WDs that exceed the Chandrasekhar mass limit {\bf ($2.1-2.8M_{\cdot}$)} \cite{AndrewHowell2006, Hicken2007, Yamanaka2009, Scalzo2010, Taubenberger2011, Silverman2011}, being termed ``super-Chandrasekhar WDs''.

WDs usually are not considered as a ``laboratory'' for strong field regimes. However, general relativistic effects have shown to be non-negligible in the massive and very magnetic WD regime \cite{Boshkayev2013a,Bera2016,Jaziel2014,Franzon2015,Otoniel2016}. Particularly, it has been shown in \cite{Bera2016} and \cite{Deu2014} that the inclusion of general relativistic effects tends to reduce the maximum mass of WDs with strong magnetic fields. Chandrasekhar and Tooper showed that instability criteria for WDs under radial oscillations in a general relativistic framework has its consequences for critical central density depending on the composition of the star \cite{Ji2013,ChandraTooper}. Moreover, in \cite{Carvalho2015,Carvalho2015a,Carvalho2017} it was shown that this critical central density yields a larger minimum radius in comparison with Newtonian gravity outcomes.

On this regard, despite the detection of gravitational wa\-ves \cite{LIGOScientificCollaborationandVirgoCollaboration2016} and innumerable other positive results, as one can check, for instance, in \cite{Will2014}, there are two cosmological phenomena, namely dark energy and dark matter, not well understood within the General Relativity (GR) framework. Such a dark sector of the universe composition is a possible indication that GR is not the ultimate theory of gravity, but a particular case of a fundamental theory.

Also, the discovery of a very massive pulsar (PSR J1614-2230) \cite{Demorest2010}, with mass $M=1.97\pm 0.04M_{\odot}$, has led to interpretative problems either on the neutron star physics or on the background theory of gravity.

In fact, compact astrophysical objects, such as black ho\-les and neutron stars, are often used as a tool to constrain extended gravity theories. For instance, in \cite{Berti2005,Moraes2014a} the bounds that could be placed on different gravitational theories using gravitational wave detection from spiraling compact binaries were investigated.

In particular, WD properties have been recently verified from extended theories of gravity, as it can be checked, for instance, in \cite{Das2015,Das2015a}.

The authors of Ref. \cite{Das2015a} have explored WD properties from an extension of GR, named $f(R)$ gravity, with $R$ being the Ricci scalar. They have shown that extended theories of gravity effects are significant in high density WDs and that the Chandrasekhar limit is not unique. By assuming $f(R)=R+\alpha R^2$, with constant $\alpha$, they have obtained super and sub-Chandrasekhar limiting mass WDs, depending on the magnitude and sign of $\alpha$, getting in touch with observations hardly explained within GR framework.

WD properties also proved to be useful to constrain modified theories of gravity. For instance, in \cite{Jain2016} the free parameters of a scalar-tensor theory of gravity was widely restricted by WD observational data.

In the present work, we are interested in analyzing WDs within the $f(R,T)$ gravity \cite{Harko2011}, with $T$ being the trace of the energy-momentum tensor. More precisely, we will investigate the hydrostatic equilibrium \cite{Tolman1939,Oppenheimer1939} of WDs in such a theory.

The $f(R,T)$ gravity has as its starting point a gravitational action which depends generally on $R$ and $T$. In this way, after the application of the variational principle, the field equations of the model are expected to present correction terms on both geometrical and material sides.

The $f(R,T)$ gravity application is motivated by its recent outcomes in different areas. It has been shown that from a minimal coupling between matter and geometry, predicted in $f(R,T)$ theories, it is possible to obtain a flat rotation curve in the halo of galaxies \cite{Zaregonbadi2016}. $f(R,T)$ models passed through solar system tests in \cite{Shabani2014}. A complete cosmological scenario was constructed from the $f(R,T)$ gravity in \cite{Moraes2016d}. In \cite{ms/2017}, a cosmological model in accordance with observations was obtained from the simplest non-minimal matter-geometry coupling within the $f(R,T)$ formalism. The validity of first and second laws of thermodynamics was discussed in \cite{Sharif2012}.

Furthermore, the hydrostatic equilibrium equation in the $f(R,T)$ gravity was originally derived in \cite{Moraes2016} and further studied in \cite{das/2016}. In \cite{sharif/2014}, the stability of collapsing spherical body coupled with isotropic matter was investigated. In \cite{noureen/2015}, the collapse equation in $f(R,T)$ gravity was derived from the perturbation scheme application. In \cite{noureen/2015b}, the instability range of the $f(R,T)$ gravity for an anisotropic background constrained by zero expansions has been developed. Moreover, the evolution of a spherical star by employing a perturbation scheme was explored in \cite{noureen/2015c,Zubair2015}.

As it shall be outlined below, the $f(R,T)$ gravity may be an important tool to study WD macroscopical properties. It also can present some advantages when compared to $f(R)$ gravity predictions for such objects, as we will show.

\section{A brief review of the $f(R,T)$ gravity formalism}\label{sec:frt}

Proposed by Harko et al. \cite{Harko2011}, the $f(R,T)$ gravity is a generalization of the $f(R)$ theories (check, for instance, \cite{Nojiri2011}). Its gravitational action depends on an arbitrary function of both the Ricci scalar $R$ and the trace of the energy-momentum tensor $T$. The dependence on $T$ is inspired by the consideration of quantum effects.

The $f(R,T)$ action reads \cite{Harko2011}
\begin{equation}\label{acao}
  \mathcal{S}=\int d^{4}x\sqrt{-g}\left[  \frac{f(R,T)}{16\pi}+\mathcal{L}_{m}\right].
\end{equation}
In (\ref{acao}), $f(R,T)$ is the general function of $R$ and $T$, $\mathcal{L}_{m}$ is the matter Lagrangian density and $g$ is the determinant of the metric tensor $g_{\mu\nu}$. Throughout this work, it will be considered the metric signature $-2$ and $c=1=G$.

The field equations of the theory are obtained by varying the action with respect to the metric $g_{\mu\nu}$, yielding \cite{Harko2011}
\begin{multline}\label{eqcampo}
  f_{R}(R,T)R_{\mu\nu}-\frac{1}{2}f(R,T)g_{\mu\nu}+(g_{\mu\nu}\Box-\nabla_{\mu}\nabla_{\nu})f_{R}(R,T)=\\8\pi T_{\mu\nu}-f_{T}(R,T)(T_{\mu\nu}+\Theta_{\mu\nu}),
\end{multline}
where
\begin{eqnarray}
  f_{R}(R,T)\equiv\frac{\partial f(R,T)}{\partial R},\;\;f_{T}(R,T)\equiv\frac{\partial f(R,T)}{\partial T},
\\
\Theta_{\mu\nu}\equiv g^{\alpha\beta}\frac{\delta T_{\alpha\beta}}{\delta g^{\mu\nu}},\;\; T_{\mu\nu}=g_{\mu\nu}\mathcal{L}_{m}-2\frac{\partial\mathcal{L}_{m}}{\partial g^{\mu\nu}}.
\end{eqnarray}
Still in Equation \eqref{eqcampo} above, $R_{\mu\nu}$ represents the Ricci tensor, $\Box$=$\nabla^{\mu}\nabla_{\mu}$ is the D'Alembertian and $\nabla_{\mu}$ is the covariant derivative.

From the covariant derivative of the field equation (\ref{eqcampo}), one obtains \cite{Alvarenga2013a,BarrientosO.2014,Moraes2016}
\begin{multline}
  \label{divtensor}
  \nabla^{\mu}T_{\mu\nu}=\frac{f_{T}(R,T)}{8\pi-f_{T}(R,T)} \times \\\left[(T_{\mu\nu}+\Theta_{\mu\nu})\nabla^{\mu}\ln f_{T}(R,T)-\frac{1}{2}g_{\mu\nu}\nabla^{\mu}T+\nabla^{\mu}\Theta_{\mu\nu}\right].
\end{multline}

We will consider the energy-momentum tensor of a perfect fluid, such that
\begin{equation}
T_{\mu\nu}=(p+\rho)u_{\mu}u_{\nu}-pg_{\mu\nu},
\end{equation}
where $p$ and $\rho$ represent the pressure and the energy density of the fluid, respectively, and $u_{\mu}$ is the four velocity of the fluid, with $u_{\mu}u^{\mu}=1$ and $u^{\mu}\nabla_{\nu}u_{\mu}=0$. The energy-momentum tensor and the conditions aforementioned imply that
\begin{eqnarray}
\mathcal{L}_{m}=-p,\\
\Theta_{\mu\nu}=-pg_{\mu\nu}-2T_{\mu\nu}.
\end{eqnarray}

In order to obtain exact solutions in the $f(R,T)$ theory, it is necessary to consider a specific form for the function $f(R,T)$. Following a previous work \cite{Moraes2016}, we will consider the functional form $f(R,T)=R+2\,f(T)$ with $f(T)$
$=\lambda T$ and $\lambda$ a constant. Such a functional form has been broadly applied in $f(R,T)$ models \cite{Alvarenga2013, Moraes2014, Moraes2015, Shamir2015,Moraes2016a} and allows the recovering of GR by simply taking $\lambda=0$.

By considering $f(R,T)=R+2\lambda T$ in Eqs.~\eqref{eqcampo} and \eqref{divtensor}, it follows that
\begin{eqnarray}
&&G_{\mu\nu}=8\pi T_{\mu\nu}+\lambda [Tg_{\mu\nu} +2(T_{\mu\nu}+pg_{\mu\nu})],\label{eqcampo2}
\\
&&\nabla^{\mu}T_{\mu\nu}=-\frac{2\lambda}{8\pi+2\lambda} \left[\nabla^{\mu}(pg_{\mu\nu})+\frac{1}{2}g_{\mu\nu}\nabla^{\mu}T\right],\label{divtensor2}
\end{eqnarray}
with $G_{\mu\nu}$ in Eq.~\eqref{eqcampo2} representing the usual Einstein tensor. 

\section{Stellar structure equations in $f(R,T)$ gravity}

The line element used to describe spherical objects follows the form:
\begin{equation}\label{line_element}
ds^2=e^{a(r)}dt^2-e^{b(r)}dr^2-r^2(d\theta^2+\sin^2\theta d\phi^2),
\end{equation}
where $(t,r,\theta,\phi)$ are the Schwarzschild-like coordinates and the exponents $a(r)$ and $b(r)$ are functions of the radial coordinate $r$.

Considering the space-time metric \eqref{line_element} in the field equation \eqref{eqcampo2} we obtain:
\begin{eqnarray}
G_{0}^{0} = \frac{e^{-b}}{r^2}\left(b'r+e^{b}-1\right) = 8\pi\rho + \lambda(3\rho - p),\label{eq:e:11}
\\
G_{1}^{1} = -\frac{e^{-b}}{r^2}\left(a'r-e^{b}+1\right) = -8\pi p + \lambda(\rho - 3p),\label{eq:e:22}
\\
G_{2}^{2}  =
G_{3}^{3} = \frac{e^{-b}}{4r}\left( \left(a'b'-2a''-a'^2 \right)r + 2(b' - a') \right)\nonumber \\ 
= -8\pi p + \lambda(\rho - 3p),\label{eq:e:33}
\end{eqnarray}
in which primes ($'$) indicate derivatives with respect to $r$.

Now, we introduce a new function $m(r)$ which depends on the radial coordinate only, in such a form that
\begin{equation}\label{radialmetric}
e^{-b}=1-\frac{2m}{r}.
\end{equation}
By replacing it in Eq.~\eqref{eq:e:11} yields
\begin{equation}\label{cmass}
\frac{dm}{dr}= 4\pi \rho r^2 +\frac{\lambda}{2}(3\rho-p)r^2,
\end{equation}
for which the function $m=m(r)$ represents the gravitational mass enclosed in a surface of radius $r$ according to the $f(R,T)$ gravity.

An additional equation is derived from  \eqref{divtensor2} and reads
\begin{equation}\label{eq:gradp:2}
\frac{dp}{dr}+\left(\rho+p\right)\frac{a'}{2} = -\frac{\lambda}{8\pi+2\lambda}(p'-\rho').
\end{equation}

Considering the relation $\rho=\rho(p)$ and Eqs.~\eqref{eq:e:22} and \eqref{radialmetric} in \eqref{eq:gradp:2}, the hydrostatic equilibrium equation for the $f(R,T)=R+2\lambda T$ gravity is obtained as
\begin{multline}\label{mtov}
\frac{dp}{dr}=-(p+\rho) \left[4\pi pr+\frac{m}{r^2}-\frac{\lambda(\rho-3p)r}{2}\right]\left(1-\frac{2m}{r}\right)^{-1}\times \\ \left[1+\frac{\lambda}{8\pi+2\lambda}\left(1-\frac{d\rho}{dp}\right)\right]^{-1}.
\end{multline}

It is quite simple to recover the usual TOV equation \cite{Tolman1939, Oppenheimer1939} in \eqref{mtov} by making $\lambda=0$. 

We remark that stellar equilibrium configurations are found only for:
\begin{equation}\label{last_term}
\frac{\lambda}{8\pi+2\lambda}\left(1-\frac{d\rho}{dp}\right)>-1.
\end{equation}
If Eq.\eqref{last_term} is not satisfied, the sign of the pressure gradient is changed, what makes the pressure to grow up from the center of the star to its surface, instead of decreasing, which is necessary for the star hydrostatic equilibrium. Since the sound velocity $v_s^2=dp/d\rho$ is in the interval $0<dp/d\rho<1$ and for the WD equation of state (EoS), where the electron degeneracy pressure is very small compared to the energy density, due to the very large ion contribution, $d\rho/dp$ becomes very large, and we can rewrite (\ref{last_term}) as
\begin{equation}\label{last_term2}
\frac{\lambda}{8\pi+2\lambda}<\frac{dp}{d\rho}.
\end{equation}
Considering that $dp/d\rho$ tends to zero at the surface of the WD,  we have from \eqref{last_term2} that only negative values for $\lambda$ are allowed.

\section{Numerical procedure, boundary conditions and equation of state}

By using the set of equations \eqref{eos} and \eqref{energy}, the equilibrium equations \eqref{cmass} and \eqref{mtov} will be solved numerically through the Runge-Kutta 4th-order method for diverse values of central density $\rho_c$ and $\lambda$. 

The boundary conditions in $f(R,T)$ gravity will be the same as in GR, i.e., at the center $(r=0)$ we have
\begin{equation}
  m(0)=0, \quad p(0)=p_c \quad {\rm and} \quad \rho(0)=\rho_c.
\end{equation}
The surface of the star $r=R$ is reached when the pressure vanishes, i.e., $p(R)=0$.

The EoS which describes the fluid properties inside WDs follows the model used for complete ionized atoms embedded in a relativistic Fermi gas of electrons \cite{Chandrasekhar1931, Chandrasekhar1935}:

\begin{eqnarray}
&&p(k_F) = \frac{1}{3\pi^2\hbar^3}\int_0^{k_F}\frac{k^4}{\sqrt{k^2+m_e^2}}dk,\label{eos}
\\ 
&&\rho(k_F) =\frac{1}{\pi^2\hbar^3}\int_0^{k_F}\sqrt{k^2+m_e^2}k^2dk +\frac{m_N\mu_e}{3\pi^2 \hbar^3}k_F^3,\label{energy} 
\end{eqnarray}
where the last term of the right hand side of Eq. (\ref{energy}) is the ions energy contribution, and $m_N$ represents the nucleon mass, $m_e$ the electron mass, $k_F$ is the Fermi momentum, $\hbar$ is the reduced Planck constant and $\mu_e=A/Z$ is the ratio between the nucleon number $A$ and the atomic number $Z$ for ions, such that in the present work we use $\mu_e=2$, valid for He, Ca, and O WDs. We neglected the lattice ion energy contribution that is small and responsible for a small reduction of the WD radius \cite{Boshkayev2013a}.

\section{Results}

\begin{figure}[h!]
\centering
\begin{subfigure}[b]{0.4\textwidth}
\includegraphics[width=\textwidth]{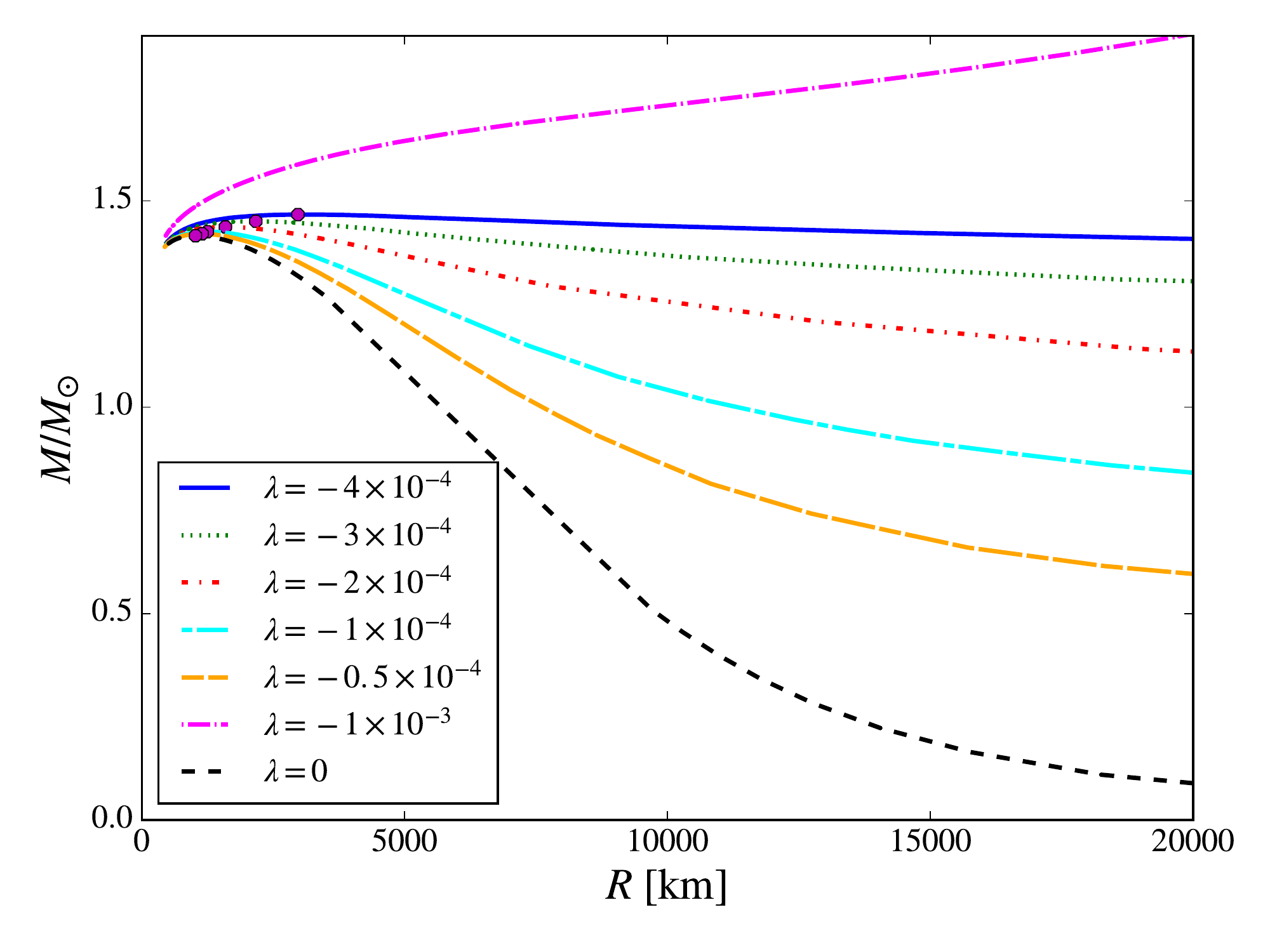}
\label{fig:total}
\end{subfigure}
\caption{\label{fig:massxradius} Total mass as a function of the total radius for different values of $\lambda$. The full magenta circles indicate the maximum mass points. 
}
\end{figure}

The mass of the WDs as a function of their total radii is shown in Fig. \ref{fig:massxradius} for six different values of $\lambda$. $\lambda=0$ recovers the GR case. 

\begin{figure}[h!]
\centering
\includegraphics[width=.4\textwidth]{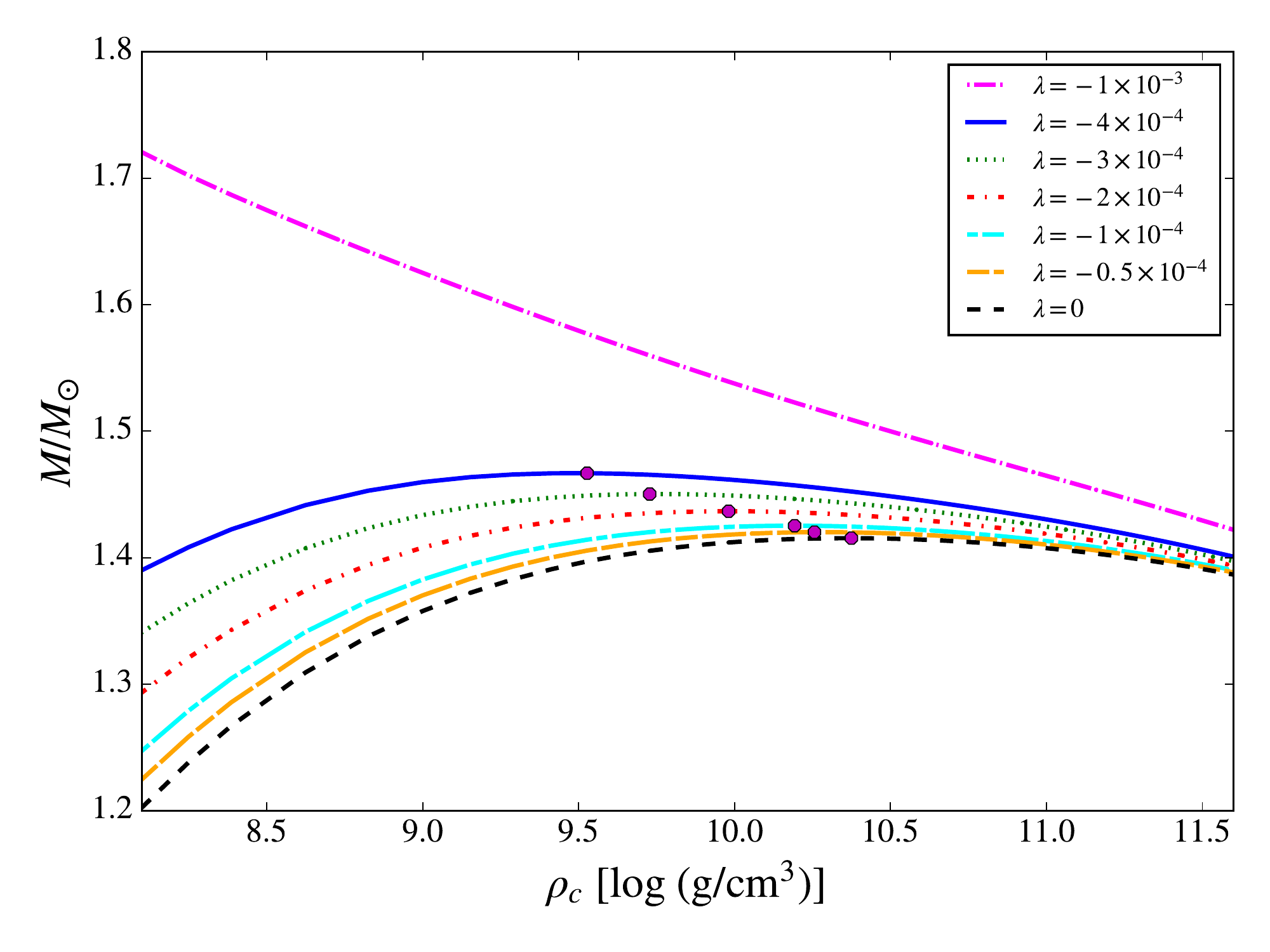}
\caption{\label{fig:massxdensity} The dependence of the total mass of the white dwarfs on central density for different values of $\lambda$.}
\end{figure}

From Figure \ref{fig:massxradius}, we note that the masses of the stars grow and  their total radii increase until attain the maximum mass point, which will be represented by full magenta circles. After that, the masses decrease with the total radii. It is important to remark that the total maximum mass grows with the decrement of $\lambda$ and the radius increases much more when we consider a fixed star mass. We also mention that the curves above tend to a plateau when $\lambda$ is $\approx -4\times 10^{-4}$. For smaller values of the parameter $\lambda$, all stars are unstable, what can be seen in Figures \ref{fig:massxradius} and \ref{fig:massxdensity} for $\lambda=-1\times 10^{-3}$, where $\partial M/\partial R<0$ and the necessary stability criterion $\partial M/\partial \rho_c>0$ are not satisfied. So, from the equilibrium configurations, the minimum value allowed for $\lambda$ is $\sim-4\times 10^{-4}$, which defines a limit for the maximum mass of the WD in the $f(R,T)$ gravity to be $\sim1.467M_{\odot}$.
\begin{figure}[h!]
\centering
\begin{subfigure}[b]{0.4\textwidth}
\includegraphics[width=\textwidth]{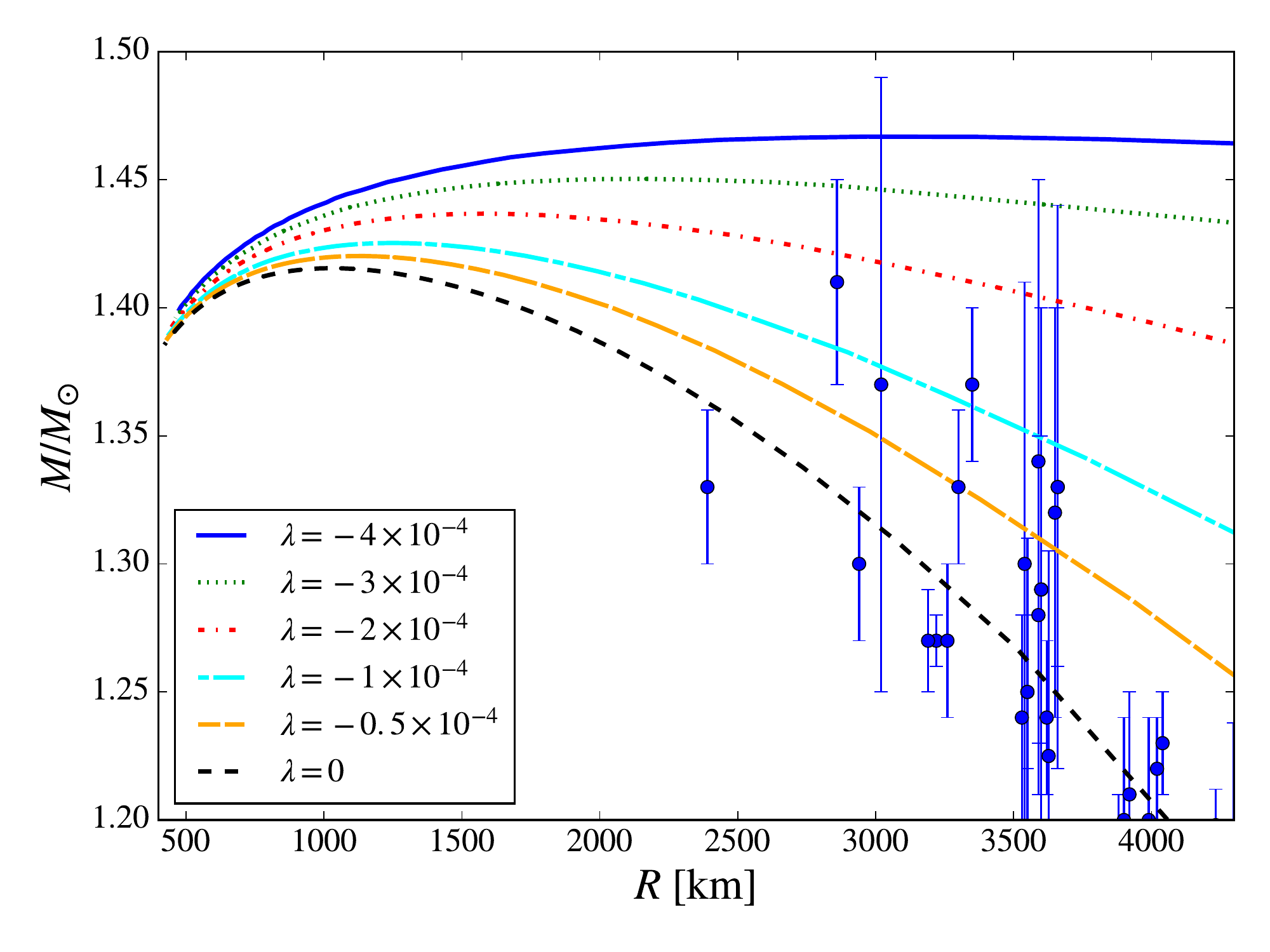}
\label{fig:total}
\end{subfigure}
\caption{\label{fig:massxradius_zoom} Mass as a function of the radius for massive WDs with different values of $\lambda$. The blue circles with error bars represent the observational data of a sample of massive WDs taken from the catalogs \cite{Vennes1997,Nalezyty2004}. 
}
\end{figure}

In addition, in Fig. \ref{fig:massxradius_zoom} we highlight the massive WDs region of Fig. \ref{fig:massxradius}, in which we have also inserted some observational data taken from the catalogs of References \cite{Vennes1997,Nalezyty2004}. It can be clearly seen from Fig. \ref{fig:massxradius_zoom} that some of the data can hardly be described purely from General Relativity, while some values of $\lambda$ can, indeed, predict the existence of massive WDs with larger radii. 

Thus, according to observations of some massive WDs, in particular the most massive, WD (1659+440J), found in \cite{Vennes1997}, the inferior limit for $\lambda$ is $\lambda_{min}\approx -3\times 10^{-4}$. We regard that this restriction is obtained by neglecting the WD data (0003+436J) with the largest error bar in Fig. \ref{fig:massxradius_zoom}. Such a constraint is more restrictive than the one obtained from Fig. \ref{fig:massxradius}, with no observational data.

\begin{figure}[ht]
  \centering
  \begin{subfigure}[b]{0.325\textwidth}
  \includegraphics[width=\textwidth]{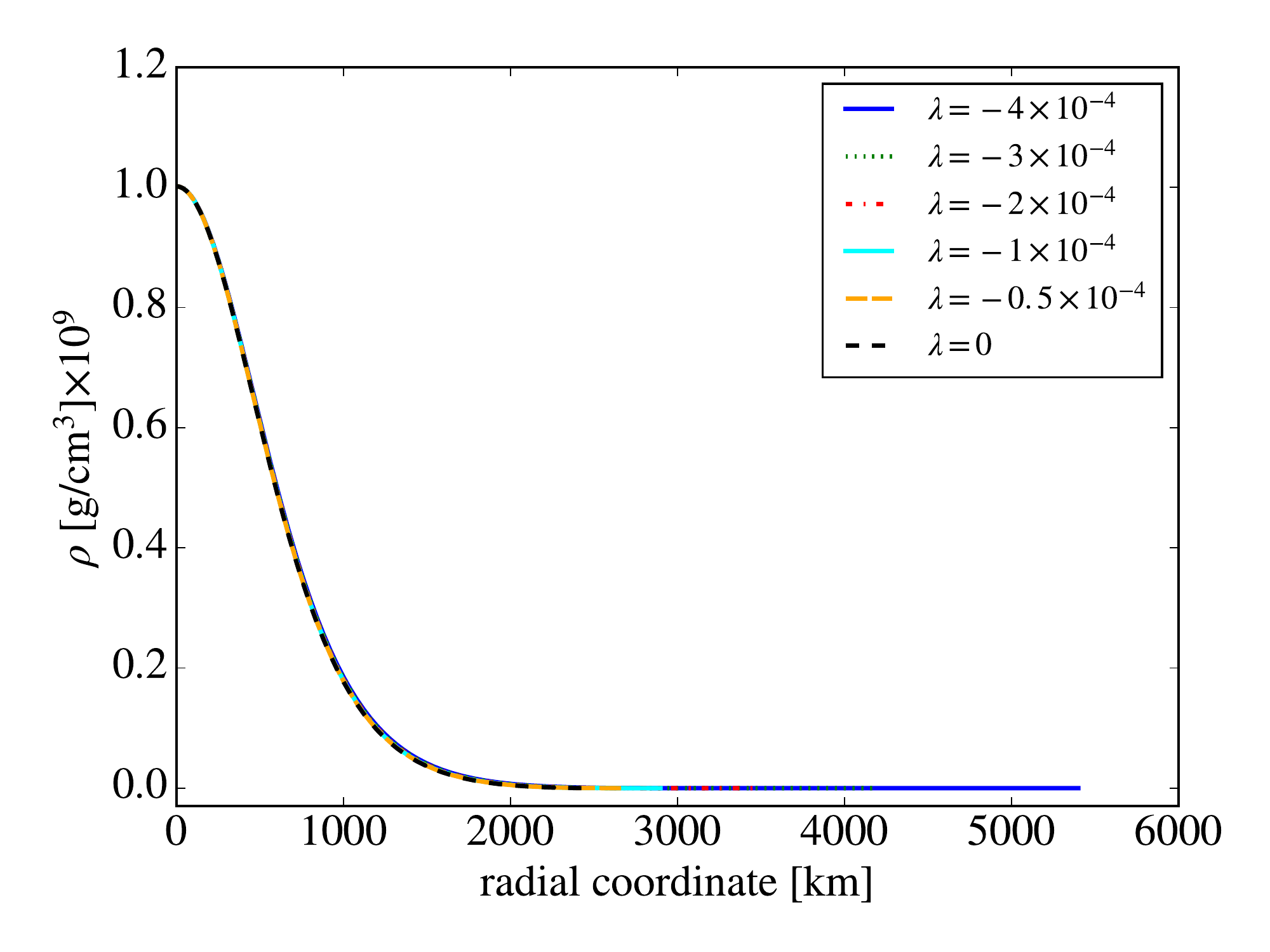}
  \label{fig:radenergy}
  \end{subfigure}
\begin{subfigure}[b]{0.325\textwidth}
  \includegraphics[width=\textwidth]{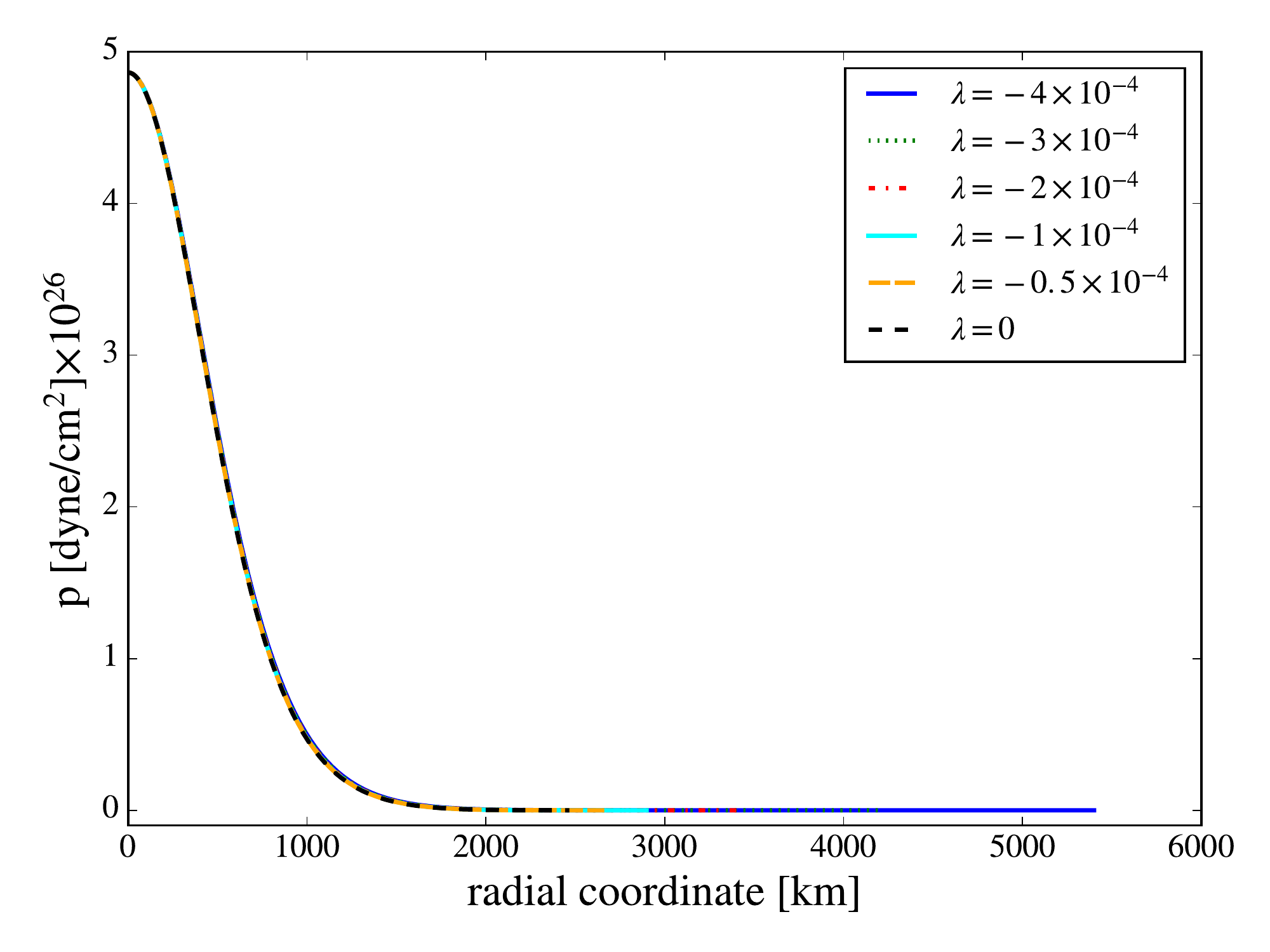}
  \label{fig:radpressure}
  \end{subfigure}
\begin{subfigure}[b]{0.325\textwidth}
  \includegraphics[width=\textwidth]{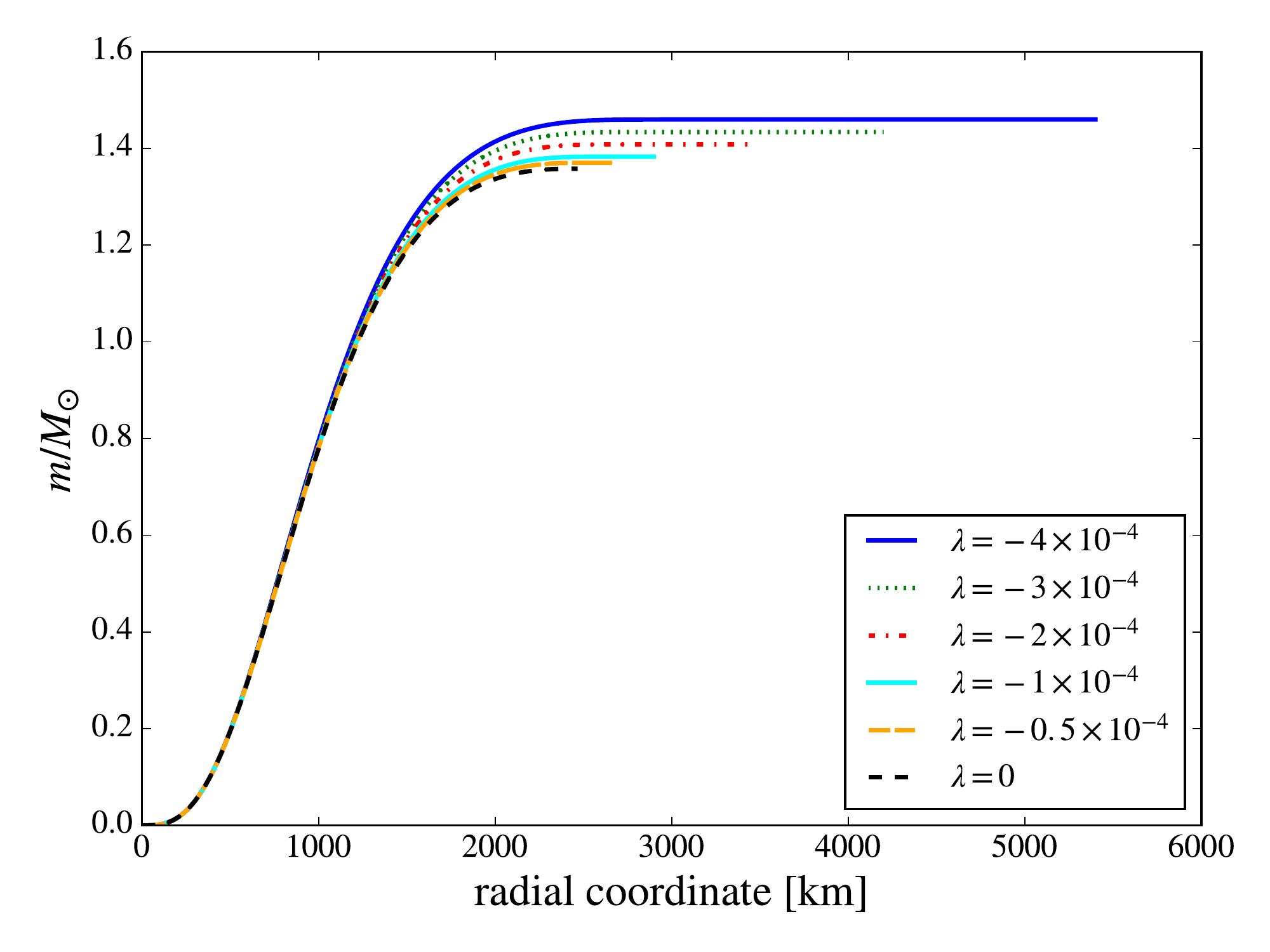}
  \label{fig:radmass}
  \end{subfigure}
  \caption{\label{fig:radialenergy} On the top panel it is presented the star energy density as a function of the radial coordinate, on the central panel we show the star pressure fluid against the radial coordinate and on the bottom panel we display the mass (in solar masses, $M_{\odot}$) inside the star versus the radial coordinate. We consider $\rho_c=10^9\ [\rm g/cm^3]$ and the displayed values of $\lambda$.}
\end{figure}

In Fig. \ref{fig:radialenergy} the energy density, fluid pressure and mass profile in the interior of the star are plotted on the top, central and bottom panels, respectively, as functions of the radial coordinate. We take into account $\rho_c=10^9\ [\rm g/cm^3]$ and different values of $\lambda$. On the top and central panels, we can observe that the energy density and the fluid pressure decrease monotonically towards the surface of the object.

On the other hand, concerning the bottom panel, it can be noted that the mass profile $m/M_{\odot}$, with $M_{\odot}$ representing the Sun's mass, grows until it reaches the surface of the star. It can also be seen that the total mass of the star increases with $\lambda$. This is due to the effect caused by the term $2\lambda T$.

Fig. \ref{fig:massxdensity} shows the behavior of the total mass against the central energy density of the stars. The values considered for the central energy density are between $1.3\times 10^8$ and $4.2\times 10^{11}\ [\rm g/cm^3]$. The upper limit is the neutron drip limit, i.e., the point where the WD turn into a neutron star. We can note that the total mass grows monotonically with central energy density until it attains a maximum value, except for $\lambda=-1\times 10^{-3}$. After that point, the stellar mass decreases with the increment of $\rho_c$ and becomes unstable.

Additionally, in Fig. \ref{fig:massxdensity}, we observe an increment of the maximum mass with $\lambda$ (see also Table \ref{tab:prop}). For example, the maximum mass value found in GR case ($\lambda=0$) is $1.417M_{\odot}$, while for $\lambda=-4\times10^{-4}$, it is $1.467M_{\odot}$.  A similar effect for $\lambda$ in the structure of the stars has been found for neutron stars and strange stars \cite{Moraes2016,Manuel2003}. Moreover, it is remarkable that for lower values of $\lambda$, the maximum mass point is reached for lower values of $\rho_c$, which can be considered an advantage of this approach when compared with $f(R)$ theory of gravity or GR outcomes, as we will argue in the next section. 

\begin{figure}[h]
\centering
\includegraphics[width=.4\textwidth]{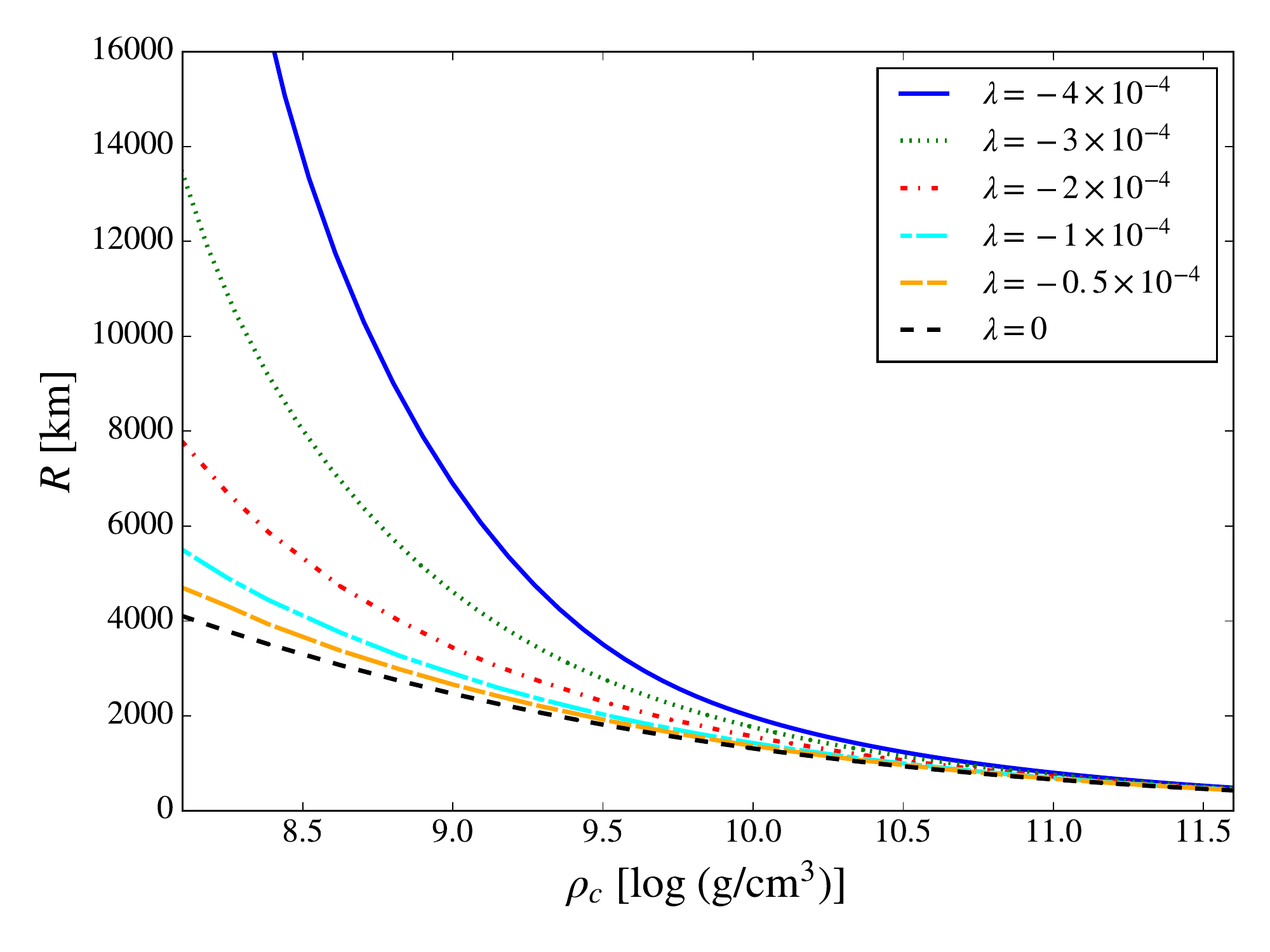}
\caption{\label{fig:cdensityradius} The total star radius versus central density for different values of $\lambda$.}
\end{figure}

In Fig. \ref{fig:cdensityradius} the dependence of the total radius with the central energy density is shown. In all cases presented, we can note that the total radius decreases when the central energy density is incremented. Larger radii are found for smaller central energy densities when $\lambda$ is decreased. This is the most important effect of $f(R,T)$ theory for massive WDs, that is, the increase of the radius, and as a consequence, the decrease of the central density, in comparison with GR and also $f(R)$ results. This is mainly due to the fact that the sound velocity becomes very small near the surface of the star, so that the term \eqref{last_term} weakens the gradient of pressure, yielding to the predicted larger radii.

In Table \ref{tab:prop} the maximum masses are presented, with their total radii and central energy densities, for each value of $\lambda$. We can see that more massive and larger WDs are found with the decrement of $\lambda$. The values of maximum masses are obtained for lower central densities (range of $\rho_c \sim 10^9-10^{10} {\rm [g/cm^3]}$) when compared with those predicted in the $f(R)$ gravity or GR scope ($\rho_c \sim 10^{11} {\rm [g/cm^3]}$) \cite{Das2015a}.

\begin{table}
\centering
\begin{tabular}{llll}
\hline \hline\
$\lambda$ & $M/M_{\odot}$ & $R\ [\rm km]$ & $\rho_c\ $[\rm g/cm$^3$]\\
\hline
$0.0\times10^{-4}$ & $1.416$ & $1021$ & $2.307\times 10^{10}$\\
$-0.5\times10^{-4}$ & $1.420$ & $1146$ & $1.803\times 10^{10}$\\
$-1.0\times10^{-4}$ & $1.425$ & $1247$ & $1.558\times 10^{10}$\\
$-2.0\times10^{-4}$ & $1.437$ & $1590$ & $9.567\times 10^{9}$\\
$-3.0\times10^{-4}$ & $1.450$ & $2168$ & $5.345\times 10^{9}$\\
$-4.0\times10^{-4}$ & $1.467$ & $2970$ & $3.366\times10^{9}$\\
\hline
\hline
\end{tabular}
\caption{The maximum masses of the white dwarfs found for each value of $\lambda$ with their respective total radii and central energy densities.}
\label{tab:prop}
\end{table}

\section{Discussion and conclusions}

In this paper we investigated the effects of an extended theory of gravity, namely $f(R,T)$ gravity, in WDs, by developing the hydrostatic equilibrium analysis for such a theory. Our main goal was to check the imprints of the extra material terms - coming from the $T-$dependence of the theory - on WD properties.


One can argue that the hydrostatic equilibrium of compact objects was already performed, originally in \cite{Moraes2016} and posteriorly in \cite{das/2016}, which is true, however those analysis have not considered the WDs EoS (\ref{eos})-(\ref{energy}) to close the system of equilibrium equations to be solved, which characterizes the path to obtain the new information content of the present paper when compared to \cite{Moraes2016,das/2016}. In this way, since recently it was shown that alternative gravity theories may contribute also to the macroscopical features of WDs (check, for instance, \cite{Das2015,Das2015a}), the present analysis is worthed.

The hydrostatic equilibrium configurations of WDs in alternative gravity theories others than $f(R,T)$ gravity can be seen in the recent literature. In \cite{Das2015} the consequences of modifications in GR were deeply analyzed in the WDs perspective. A similar approach can be seen in \cite{Das2015a}. In \cite{Jain2016} it was shown that WDs provide a unique setup to constrain Horndeski theories of gravity. In \cite{banerjee/2017}, it was explored the effects that WDs suffers when described in various modified gravity models, such as scalar-tensor-vector, Eddington inspired Born-Infeld and $f(R)$ theories of gravity. Furthermore, WDs have been used to constrain hypothetical variations on the gravitational constant \cite{althaus/2011,garcia-berro/2011,corsico/2013}.

The equilibrium configurations of WDs were analyzed for $f(R,T)=R+2\lambda T$ with different values of $\lambda$ and central energy densities. We showed that the extended theory of gravity affects the maximum mass and radius of WDs depending on the value of $\lambda$.

Since gravitational fields are smaller for WDs than for neutron stars or quarks stars, the scale parameter $\lambda$ used here is small when compared to the values used in Ref. \cite{Moraes2016}. In this way, WDs data can be used as a tool to constrain an inferior limit on $\lambda$, which is $\lambda_{min}\approx -3 \times 10^{-4}$.

The values of the parameter $\lambda$ used in the present article are clearly small when compared to those of Reference \cite{Moraes2016}, in which the hydrostatic equilibrium configurations of neutron and quark stars were calculated in $f(R,T)$ gravity. This may be due to the fact that the compactness $M/R$ of WDs is small when compared to those of neutron and quark stars. In fact, it can be seen in \cite{Moraes2016} that the values of $\lambda$ needed to get stable quark stars are greater than the values used for neutron stars, as a probable consequence of the higher compactness of quarks stars in relation to neutron stars. In this way, these analysis indicate that higher compactness objects would need higher deviations from GR. 

We found that for $\lambda=-4\times 10^{-4}$, the maximum mass of the WD is $1.47M_{\odot}$. This value is determined in a central energy density $\sim 85$\% lower and radius $\sim 110$\% greater than those values used to find the maximum mass value in the GR case ($\lambda=0$). The outcomes for the central energy density are also smaller than those obtained in $f(R)=R+\alpha R^2$ gravity, for different values of $\alpha$ \citep{Das2015a}.

We argue about the advantages of having WDs with lower central energy densities in the following. In \cite{mikheev/2016} some constraints on the central density of a WD were obtained. The authors have derived a system of equations and inequalities that allows one to determine constraints on $\rho_c$. They have found that $\rho_c\leq10^9$ $\rm g/cm^3$. Moreover, in a seminal paper by Hamada and Salpeter \cite{hamada/1961}, it was found that for the maximum masses of WDs, $\rho_c\sim10^9-10^{10}$ $\rm g/cm^3$. Recently, WD calculations in GR also showed that central energy densities are limited by nuclear fusion reactions \cite{Chamel2013,Boshkayev2013a,Otoniel2016}.  It is worth quoting that the values of the central energy densities that we have obtained for the $f(R,T)$ gravity respect these constraints. In contrast, what has been found for the central energy density of WDs in $f(R)$ gravity is $\rho_c\sim10^{11}$ $\rm g/cm^3$ \cite{Das2015a}.

As a direct extension of the present work, one can also consider quadratic terms on $T$ for the functional form of $f(R,T)$, that is, $f(R,T)=R+2\lambda T+\xi T^2$, with $\xi$ being a free parameter. Since the extra material terms seem to yield an increment on the mass of WDs, one may expect the presence of the quadratic term $T^2$ to significantly elevate the Chandrasekhar limit and predict the existence of super-Chandrasekhar WDs \cite{AndrewHowell2006, Scalzo2010}, which still require convincing physical explanation. 


Also, in a further work, the analysis presented here can be performed in models of nonminimal torsion-matter coupling, like those in References \cite{harko/2014,harko/2014b}. While the hydrostatic equilibrium of quark stars has already been performed in such models \cite{pace/2017}, the application for neutron stars and WDs still lacks.

\begin{acknowledgements}
GAC would like to thank CAPES (Coordena\c c\~ao de Aperfei\c coamento de Pessoal de N\'ivel Superior) for financial support. RVL thanks CNPq (Conselho Nacional de Desenvolvimeno Cient\'ifico e Tecnol\'ogico) process 141157/2015-1 and CAPES/PDSE/88881. 134089/2016-01 for financial support. He is also sincerely grateful to staff of ICRANet, Pescara, Italy, for kind hospitality during his visit. PHRSM would like to thank S\~ao Paulo Research Foundation (FAPESP), grant 2015/08476-0, for financial support. JDVA, RMM and MM acknowledge CAPES, CNPq and FAPESP thematic project 13/26258-4.

All computations were performed in open source software and the authors are sincerely thankful to open source community \cite{Wehbring2016,Droettboom2017,Python,Maxima}.
\end{acknowledgements}


\begin{thebibliography}{10}

\bibitem{Kepler2017} S.O. Kepler, A.D. Romero, I.~Pelisoli, G.~Ourique, arXiv:1702.01134 [astro-ph.SR]  (2017)

\bibitem{Chandrasekhar1931} S.~Chandrasekhar, The Astrophysical Journal \textbf{74}, 81 (1931).

\bibitem{Weinberg2008} S.~Weinberg, \emph{{Cosmology}} (OUP Oxford, 2008). 

\bibitem{Riess1998} A.G. Riess, A.V. Filippenko, P.~Challis, {\it et al.}, The Astronomical Journal \textbf{116}(3), 1009 (1998).

\bibitem{Perlmutter1999}
S.~Perlmutter, G.~Aldering, G.~Goldhaber, R.A. Knop, P.~Nugent, P.G. Castro,
  S.~Deustua, S.~Fabbro, {A. Goobar}, D.E. Groom, I.M. Hook, A.G. Kim, M.Y.
  Kim, J.C. Lee, N.J. Nunes, R.~Pain, C.R. Pennypacker, {R. Quimby}, C.~Lidman,
  R.S. Ellis, M.~Irwin, R.G. McMahon, P.~Ruiz-Lapuente, N.~Walton, B.~Schaefer,
  B.J. Boyle, A.V. Filippenko, T.~Matheson, A.S. Fruchter, N.~Panagia, H.J.M.
  Newberg, W.J. Couch, T.S.C. Project, The Astrophysical Journal
  \textbf{517}(2), 565 (1999).
\newblock \doi{10.1086/307221}.
\newblock \urlprefix\url{http://stacks.iop.org/0004-637X/517/i=2/a=565
  http://iopscience.iop.org/article/10.1086/307221/pdf}

\bibitem{AndrewHowell2006}
D.~Andrew~Howell, M.~Sullivan, P.E. Nugent, R.S. Ellis, A.J. Conley,
  D.~Le~Borgne, R.G. Carlberg, J.~Guy, D.~Balam, S.~Basa, D.~Fouchez, I.M.
  Hook, E.Y. Hsiao, J.D. Neill, R.~Pain, K.M. Perrett, C.J. Pritchet, Nature
  \textbf{443}(7109), 308 (2006).
\newblock \doi{10.1038/nature05103}.
\newblock \urlprefix\url{http://www.nature.com/doifinder/10.1038/nature05103}

\bibitem{Scalzo2010}
R.A. Scalzo, G.~Aldering, P.~Antilogus, C.~Aragon, S.~Bailey, C.~Baltay,
  S.~Bongard, C.~Buton, {M. Childress}, N.~Chotard, Y.~Copin, H.K. Fakhouri,
  A.~Gal-Yam, E.~Gangler, S.~Hoyer, M.~Kasliwal, S.~Loken, {P. Nugent},
  R.~Pain, E.~P{\'{e}}contal, R.~Pereira, S.~Perlmutter, D.~Rabinowitz, A.~Rau,
  G.~Rigaudier, K.~Runge, {G. Smadja}, C.~Tao, R.C. Thomas, B.~Weaver, C.~Wu,
  The Astrophysical Journal \textbf{713}(2), 1073 (2010).
\newblock \doi{10.1088/0004-637X/713/2/1073}

\bibitem{Hicken2007}
M.~Hicken, P.M. Garnavich, J.L. Prieto, S.~Blondin, D.L. DePoy, R.P. Kirshner,
  J.~Parrent, The Astrophysical Journal Letters \textbf{669}(1), L17 (2007).
\newblock \doi{10.1086/523301}

\bibitem{Yamanaka2009}
M.~Yamanaka, K.S. Kawabata, K.~Kinugasa, M.~Tanaka, A.~Imada, K.~Maeda,
  K.~Nomoto, A.~Arai, S.~Chiyonobu, {Y. Fukazawa}, O.~Hashimoto, S.~Honda,
  Y.~Ikejiri, R.~Itoh, Y.~Kamata, N.~Kawai, T.~Komatsu, K.~Konishi, {D.
  Kuroda}, H.~Miyamoto, S.~Miyazaki, O.~Nagae, H.~Nakaya, T.~Ohsugi,
  T.~Omodaka, N.~Sakai, M.~Sasada, {M. Suzuki}, H.~Taguchi, H.~Takahashi,
  H.~Tanaka, M.~Uemura, T.~Yamashita, K.~Yanagisawa, M.~Yoshida, The
  Astrophysical Journal Letters \textbf{707}(2), L118 (2009).
\newblock \doi{10.1088/0004-637X/707/2/L118}

\bibitem{Taubenberger2011}
S.~Taubenberger, S.~Benetti, M.~Childress, R.~Pakmor, S.~Hachinger, P.A.
  Mazzali, V.~Stanishev, N.~Elias-Rosa, I.~Agnoletto, F.~Bufano, M.~Ergon,
  A.~Harutyunyan, C.~Inserra, E.~Kankare, M.~Kromer, H.~Navasardyan,
  J.~Nicolas, A.~Pastorello, E.~Prosperi, F.~Salgado, J.~Sollerman,
  M.~Stritzinger, M.~Turatto, S.~Valenti, W.~Hillebrandt, Monthly Notices of
  the Royal Astronomical Society \textbf{412}(4), 2735 (2011).
\newblock \doi{10.1111/j.1365-2966.2010.18107.x}

\bibitem{Silverman2011}
J.M. Silverman, M.~Ganeshalingam, W.~Li, A.V. Filippenko, A.A. Miller,
  D.~Poznanski, Monthly Notices of the Royal Astronomical Society
  \textbf{410}(1), 585 (2011).
\newblock \doi{10.1111/j.1365-2966.2010.17474.x}

\bibitem{Moll2014}
R.~Moll, C.~Raskin, D.~Kasen, S.E. Woosley, The Astrophysical Journal
  \textbf{785}(2), 105 (2014).
\newblock \doi{10.1088/0004-637X/785/2/105}.
\newblock \urlprefix\url{http://stacks.iop.org/0004-637X/785/i=2/a=105
  http://iopscience.iop.org/article/10.1088/0004-637X/785/2/105/meta;jsessionid=ECBEF9C5A4F8638AAA3BA5FEAADAD38D.c2.iopscience.cld.iop.org/pdf}

\bibitem{Ji2013}
S.~Ji, R.T. Fisher, E.~Garc{\'{i}}a-Berro, P.~Tzeferacos, G.~Jordan, D.~Lee,
  {Pablo Lor{\'{e}}n-Aguilar}, P.~Cremer, J.~Behrends, The Astrophysical
  Journal \textbf{773}(2), 136 (2013).
\newblock \doi{10.1088/0004-637X/773/2/136}.
\newblock \urlprefix\url{http://stacks.iop.org/0004-637X/773/i=2/a=136
  http://iopscience.iop.org/article/10.1088/0004-637X/773/2/136/meta;jsessionid=FF4759FA41D2FF885274AFDEDA059011.c1.iopscience.cld.iop.org/pdf}

\bibitem{Rossum2016}
D.R.v. Rossum, R.~Kashyap, R.~Fisher, R.T. Wollaeger, E.~Garc{\'{i}}a-Berro,
  {Gabriela Aznar-Sigu{\'{a}}n}, S.~Ji, P.~Lor{\'{e}}n-Aguilar, The
  Astrophysical Journal \textbf{827}(2), 128 (2016).
\newblock \doi{10.3847/0004-637X/827/2/128}.
\newblock \urlprefix\url{http://stacks.iop.org/0004-637X/827/i=2/a=128
  http://iopscience.iop.org/article/10.3847/0004-637X/827/2/128/meta;jsessionid=18D05A11E77E67E9963D6FE7B014A54C.ip-10-40-1-105/pdf}

\bibitem{Das2012}
U.~Das, B.~Mukhopadhyay, Physical Review D \textbf{86}(4), 042001 (2012).
\newblock \doi{10.1103/PhysRevD.86.042001}

\bibitem{Das2012a}
U.~DAS, B.~MUKHOPADHYAY, International Journal of Modern Physics D
  \textbf{21}(11), 1242001 (2012).
\newblock \doi{10.1142/S0218271812420011}.
\newblock
  \urlprefix\url{http://www.worldscientific.com/doi/abs/10.1142/S0218271812420011}

\bibitem{Das2013}
U.~Das, B.~Mukhopadhyay, Physical Review Letters \textbf{110}(7), 071102
  (2013).
\newblock \doi{10.1103/PhysRevLett.110.071102}

\bibitem{Chamel2013}
N.~Chamel, A.F. Fantina, P.J. Davis, Physical Review D \textbf{88}(8), 081301
  (2013).
\newblock \doi{10.1103/PhysRevD.88.081301}

\bibitem{Dong2014}
J.M. Dong, W.~Zuo, P.~Yin, J.Z. Gu, Physical Review Letters \textbf{112}(3),
  039001 (2014).
\newblock \doi{10.1103/PhysRevLett.112.039001}.
\newblock
  \urlprefix\url{https://link.aps.org/doi/10.1103/PhysRevLett.112.039001
  https://journals.aps.org/prl/abstract/10.1103/PhysRevLett.112.039001}

\bibitem{Coelho2014b}
J.G. Coelho, R.M. Marinho, M.~Malheiro, R.~Negreiros, D.L. C{\'{a}}ceres, J.A.
  Rueda, R.~Ruffini, The Astrophysical Journal \textbf{794}(1), 86 (2014).
\newblock \doi{10.1088/0004-637X/794/1/86}

\bibitem{Nityananda2015}
R.~Nityananda, S.~Konar, Physical Review D \textbf{91}(2), 028301 (2015).
\newblock \doi{10.1103/PhysRevD.91.028301}

\bibitem{Liu2014}
H.~Liu, X.~Zhang, D.~Wen, Physical Review D \textbf{89}(10), 104043 (2014).
\newblock \doi{10.1103/PhysRevD.89.104043}.
\newblock \urlprefix\url{https://link.aps.org/doi/10.1103/PhysRevD.89.104043}

\bibitem{Das2014}
U.~Das, B.~Mukhopadhyay, Journal of Cosmology and Astroparticle Physics
  \textbf{2014}(06), 050 (2014).
\newblock \doi{10.1088/1475-7516/2014/06/050}

\bibitem{Boshkayev2013a}
K.~Boshkayev, J.A. Rueda, R.~Ruffini, I.~Siutsou, The Astrophysical Journal
  \textbf{762}(2), 117 (2013).
\newblock \doi{10.1088/0004-637X/762/2/117}

\bibitem{Franzon2015}
B.~Franzon, S.~Schramm, Physical Review D \textbf{92}(8), 083006 (2015).
\newblock \doi{10.1103/PhysRevD.92.083006}

\bibitem{Subramanian2015}
S.~Subramanian, B.~Mukhopadhyay, Monthly Notices of the Royal Astronomical
  Society \textbf{454}(1), 752 (2015).
\newblock \doi{10.1093/mnras/stv1983}

\bibitem{Carvalho2015}
G.A. Carvalho, R.M.M. Jr, M.~Malheiro, Journal of Physics: Conference Series
  \textbf{630}(1), 012058 (2015).
\newblock \doi{10.1088/1742-6596/630/1/012058}

\bibitem{Carvalho2015a}
G.~Carvalho, R.~Marinho, M.~Malheiro, in \emph{THE SECOND ICRANet C{\'{E}}SAR
  LATTES MEETING: Supernovae, Neutron Stars and Black Holes}, vol. 1693 (AIP
  Publishing, 2015), vol. 1693, p. 030004.
\newblock \doi{10.1063/1.4937187}

\bibitem{Bera2016}
P.~Bera, D.~Bhattacharya, Monthly Notices of the Royal Astronomical Society
  \textbf{456}(3), 3375 (2016).
\newblock \doi{10.1093/mnras/stv2823}.
\newblock \urlprefix\url{http://arxiv.org/abs/1508.05521
  http://www.arxiv.org/pdf/1508.05521.pdf
  https://academic.oup.com/mnras/article-lookup/doi/10.1093/mnras/stv2823}

\bibitem{Otoniel2016}
E.~Otoniel, B.~Franzon, M.~Malheiro, S.~Schramm, F.~Weber, arXiv:1609.05994
  [astro-ph.SR]  (2016).
\newblock \urlprefix\url{http://arxiv.org/abs/1609.05994}

\bibitem{Lynden-Bell1967}
D.~Lynden-Bell, J.P. Ostriker, Monthly Notices of the Royal Astronomical
  Society \textbf{136}, 293 (1967).
\newblock \doi{10.1093/mnras/136.3.293}.
\newblock \urlprefix\url{http://adsabs.harvard.edu/abs/1967MNRAS.136..293L
  http://adsabs.harvard.edu/cgi-bin/nph-data_query?bibcode=1967MNRAS.136..293L&link_type=ARTICLE}

\bibitem{Ostriker1968a}
J.P. Ostriker, J.W.K. Mark, The Astrophysical Journal \textbf{151}, 1075
  (1968).
\newblock \doi{10.1086/149506}.
\newblock \urlprefix\url{http://adsabs.harvard.edu/abs/1968ApJ...151.1075O
  http://adsabs.harvard.edu/cgi-bin/nph-data_query?bibcode=1968ApJ...151.1075O&link_type=ARTICLE}

\bibitem{Ostriker1968c}
J.P. Ostriker, P.~Bodenheimer, The Astrophysical Journal \textbf{151}, 1089
  (1968).
\newblock \doi{10.1086/149507}.
\newblock \urlprefix\url{http://adsabs.harvard.edu/abs/1968ApJ...151.1089O
  http://adsabs.harvard.edu/cgi-bin/nph-data_query?bibcode=1968ApJ...151.1089O&link_type=ARTICLE}

\bibitem{Ostriker1968}
J.P. Ostriker, F.D.A. Hartwick, The Astrophysical Journal \textbf{153}, 797
  (1968).
\newblock \doi{10.1086/149706}

\bibitem{Ostriker1969}
J.P. Ostriker, J.L. Tassoul, The Astrophysical Journal \textbf{155}, 987
  (1969).
\newblock \doi{10.1086/149927}.
\newblock \urlprefix\url{http://adsabs.harvard.edu/abs/1969ApJ...155..987O
  http://adsabs.harvard.edu/cgi-bin/nph-data_query?bibcode=1969ApJ...155..987O&link_type=ARTICLE}

\bibitem{Paczynski1990}
B.~Paczynski, The Astrophysical Journal Letters \textbf{365}, L9 (1990).
\newblock \doi{10.1086/185876}

\bibitem{Usov1993}
V.V. Usov, The Astrophysical Journal \textbf{410}, 761 (1993).
\newblock \doi{10.1086/172792}

\bibitem{Usov1994}
V.V. Usov, The Astrophysical Journal \textbf{427}, 984 (1993).
\newblock \doi{10.1086/174205}.
\newblock \urlprefix\url{http://adsabs.harvard.edu/abs/1994ApJ...427..984U
  http://adsabs.harvard.edu/doi/10.1086/174205
  http://arxiv.org/abs/astro-ph/9312061 http://dx.doi.org/10.1086/174205}

\bibitem{Malheiro2012}
M.~Malheiro, J.A. Rueda, R.~Ruffini, Publications of the Astronomical Society
  of Japan \textbf{64}(3), 56 (2011).
\newblock \doi{10.1093/pasj/64.3.56}.
\newblock \urlprefix\url{http://arxiv.org/abs/1102.0653
  http://www.arxiv.org/pdf/1102.0653.pdf
  http://pasj.oxfordjournals.org/content/64/3/56.short
  https://academic.oup.com/pasj/article-lookup/doi/10.1093/pasj/64.3.56
  http://dx.doi.org/10.1093/pasj/64.3.56}

\bibitem{Coelho2014c}
J.G. Coelho, M.~Malheiro, Publications of the Astronomical Society of Japan
  \textbf{66}(1), 14 (2014).
\newblock \doi{10.1093/pasj/pst014}.
\newblock
  \urlprefix\url{https://academic.oup.com/pasj/article-lookup/doi/10.1093/pasj/pst014}

\bibitem{Lobato2015}
R.V. Lobato, J.~Coelho, M.~Malheiro, Journal of Physics: Conference Series
  \textbf{630}(1), 012015 (2015).
\newblock \doi{10.1088/1742-6596/630/1/012015}.
\newblock
  \urlprefix\url{http://stacks.iop.org/1742-6596/630/i=1/a=012015?key=crossref.231be24f8f660aaa2c2f9146bb62d028}

\bibitem{Lobato2015a}
R.V. Lobato, J.G. Coelho, M.~Malheiro, in \emph{THE SECOND ICRANET C{\'{E}}SAR
  LATTES MEETING: Supernovae, Neutron Stars and Black Holes} (AIP Publishing,
  Rio de Janeiro - Niter{\'{o}}i - Jo{\~{a}}o Pessoa - Recife - Fortaleza,
  2015), p. 030003.
\newblock \doi{10.1063/1.4937186}.
\newblock \urlprefix\url{http://aip.scitation.org/doi/abs/10.1063/1.4937186}

\bibitem{Lobato2016}
R.V. Lobato, M.~Malheiro, J.G. Coelho, International Journal of Modern Physics
  D \textbf{25}(09), 1641025 (2016).
\newblock \doi{10.1142/S021827181641025X}.
\newblock \urlprefix\url{http://arxiv.org/abs/1603.00870
  http://dx.doi.org/10.1142/S021827181641025X
  http://www.worldscientific.com/doi/abs/10.1142/S021827181641025X}

\bibitem{Lobato2016b}
R.V. Lobato, M.~Malheiro, Journal of Physics: Conference Series \textbf{706},
  052032 (2016).
\newblock \doi{10.1088/1742-6596/706/5/052032}

\bibitem{Mukhopadhyay2016}
B.~Mukhopadhyay, A.~Rao, Journal of Cosmology and Astroparticle Physics
  \textbf{2016}(05), 007 (2016).
\newblock \doi{10.1088/1475-7516/2016/05/007}.
\newblock
  \urlprefix\url{http://arxiv.org/abs/1603.00141%5Cnhttp://stacks.iop.org/1475-7516/2016/i=05/a=007?key=crossref.f535ad740862e0c75adede030b6e9558}

\bibitem{Lobato2017}
R.V. Lobato, J.G. Coelho, M.~Malheiro,   (2017).
\newblock \urlprefix\url{http://arxiv.org/abs/1703.06208}

\bibitem{Castanheira2013}
B.G. Castanheira, S.O. Kepler, S.J. Kleinman, A.~Nitta, L.~Fraga, Monthly
  Notices of the Royal Astronomical Society \textbf{430}(1), 50 (2013).
\newblock \doi{10.1093/mnras/sts474}

\bibitem{Kepler2013}
S.O. Kepler, I.~Pelisoli, S.~Jordan, S.J. Kleinman, D.~Koester, B.~K, V.~P,
  B.G. Castanheira, A.~Nitta, J.E.S. Costa, D.E. Winget, A.~Kanaan, L.~Fraga,
  Monthly Notices of the Royal Astronomical Society \textbf{429}(4), 2934
  (2013).
\newblock \doi{10.1093/mnras/sts522}

\bibitem{Marsh2016}
T.R. Marsh, B.T. G{\"{a}}nsicke, S.~H{\"{u}}mmerich, F.J. Hambsch, K.~Bernhard,
  C.~Lloyd, E.~Breedt, E.R. Stanway, D.T. Steeghs, S.G. Parsons, O.~Toloza,
  M.R. Schreiber, P.G. Jonker, J.~van Roestel, T.~Kupfer, A.F. Pala, V.S.
  Dhillon, L.K. Hardy, S.P. Littlefair, A.~Aungwerojwit, S.~Arjyotha,
  D.~Koester, J.J. Bochinski, C.A. Haswell, P.~Frank, P.J. Wheatley, Nature pp.
  1--15 (2016).
\newblock \doi{10.1038/nature18620}

\bibitem{Buckley2017}
D.A.H. Buckley, P.J. Meintjes, S.B. Potter, T.R. Marsh, B.T. G{\"{a}}nsicke,
  Nature Astronomy \textbf{1}(2), 0029 (2017).
\newblock \doi{10.1038/s41550-016-0029}

\bibitem{Das2015a}
U.~Das, B.~Mukhopadhyay, Journal of Cosmology and Astroparticle Physics
  \textbf{2015}(05), 045 (2015).
\newblock \doi{10.1088/1475-7516/2015/05/045}.
\newblock \urlprefix\url{http://stacks.iop.org/1475-7516/2015/i=05/a=045
  http://iopscience.iop.org/article/10.1088/1475-7516/2015/05/045/pdf
  http://stacks.iop.org/1475-7516/2015/i=05/a=045?key=crossref.352fee117fa89bce140f080283c5dd95}

\bibitem{LIGOScientificCollaborationandVirgoCollaboration2016}
{LIGO Scientific Collaboration and Virgo Collaboration}, B.P. Abbott,
  R.~Abbott, T.D. Abbott, M.R. Abernathy, F.~Acernese, K.~Ackley, C.~Adams,
  T.~Adams, P.~Addesso, R.X. Adhikari, V.B. Adya, C.~Affeldt, M.~Agathos,
  K.~Agatsuma, N.~Aggarwal, O.D. Aguiar, L.~Aiello, A.~Ain, P.~Ajith, B.~Allen,
  A.~Allocca, P.A. Altin, S.B. Anderson, W.G. Anderson, K.~Arai, M.A. Arain,
  M.C. Araya, C.C. Arceneaux, J.S. Areeda, N.~Arnaud, K.G. Arun, S.~Ascenzi,
  G.~Ashton, M.~Ast, S.M. Aston, P.~Astone, P.~Aufmuth, C.~Aulbert, S.~Babak,
  P.~Bacon, M.K.M. Bader, P.T. Baker, F.~Baldaccini, G.~Ballardin, S.W.
  Ballmer, J.C. Barayoga, S.E. Barclay, B.C. Barish, D.~Barker, F.~Barone,
  B.~Barr, L.~Barsotti, M.~Barsuglia, D.~Barta, J.~Bartlett, M.A. Barton,
  I.~Bartos, R.~Bassiri, A.~Basti, J.C. Batch, C.~Baune, V.~Bavigadda,
  M.~Bazzan, B.~Behnke, M.~Bejger, C.~Belczynski, A.S. Bell, C.J. Bell, B.K.
  Berger, J.~Bergman, G.~Bergmann, C.P.L. Berry, D.~Bersanetti, A.~Bertolini,
  J.~Betzwieser, S.~Bhagwat, R.~Bhandare, I.A. Bilenko, G.~Billingsley,
  J.~Birch, R.~Birney, O.~Birnholtz, S.~Biscans, A.~Bisht, M.~Bitossi,
  C.~Biwer, M.A. Bizouard, J.K. Blackburn, C.D. Blair, D.G. Blair, R.M. Blair,
  S.~Bloemen, O.~Bock, T.P. Bodiya, M.~Boer, G.~Bogaert, C.~Bogan, A.~Bohe,
  P.~Bojtos, C.~Bond, F.~Bondu, R.~Bonnand, B.A. Boom, R.~Bork, V.~Boschi,
  S.~Bose, Y.~Bouffanais, A.~Bozzi, C.~Bradaschia, P.R. Brady, V.B. Braginsky,
  M.~Branchesi, J.E. Brau, T.~Briant, A.~Brillet, M.~Brinkmann, V.~Brisson,
  P.~Brockill, A.F. Brooks, D.A. Brown, D.D. Brown, N.M. Brown, C.C. Buchanan,
  A.~Buikema, T.~Bulik, H.J. Bulten, A.~Buonanno, D.~Buskulic, C.~Buy, R.L.
  Byer, M.~Cabero, L.~Cadonati, G.~Cagnoli, C.~Cahillane, J.C. Bustillo,
  T.~Callister, E.~Calloni, J.B. Camp, K.C. Cannon, J.~Cao, C.D. Capano,
  E.~Capocasa, F.~Carbognani, S.~Caride, J.C. Diaz, C.~Casentini, S.~Caudill,
  M.~Cavagli{\`{a}}, F.~Cavalier, R.~Cavalieri, G.~Cella, C.B. Cepeda, L.C.
  Baiardi, G.~Cerretani, E.~Cesarini, R.~Chakraborty, T.~Chalermsongsak, S.J.
  Chamberlin, M.~Chan, S.~Chao, P.~Charlton, E.~Chassande-Mottin, H.Y. Chen,
  Y.~Chen, C.~Cheng, A.~Chincarini, A.~Chiummo, H.S. Cho, M.~Cho, J.H. Chow,
  N.~Christensen, Q.~Chu, S.~Chua, S.~Chung, G.~Ciani, F.~Clara, J.A. Clark,
  F.~Cleva, E.~Coccia, P.F. Cohadon, A.~Colla, C.G. Collette, L.~Cominsky,
  M.~Constancio, A.~Conte, L.~Conti, D.~Cook, T.R. Corbitt, N.~Cornish,
  A.~Corsi, S.~Cortese, C.A. Costa, M.W. Coughlin, S.B. Coughlin, J.P. Coulon,
  S.T. Countryman, P.~Couvares, E.E. Cowan, D.M. Coward, M.J. Cowart, D.C.
  Coyne, R.~Coyne, K.~Craig, J.D.E. Creighton, T.D. Creighton, J.~Cripe, S.G.
  Crowder, A.M. Cruise, A.~Cumming, L.~Cunningham, E.~Cuoco, T.D. Canton, S.L.
  Danilishin, S.~D’Antonio, K.~Danzmann, N.S. Darman, C.F. Da~Silva~Costa,
  V.~Dattilo, I.~Dave, H.P. Daveloza, M.~Davier, G.S. Davies, E.J. Daw, R.~Day,
  S.~De, D.~DeBra, G.~Debreczeni, J.~Degallaix, M.~De~Laurentis,
  S.~Del{\'{e}}glise, W.~Del~Pozzo, T.~Denker, T.~Dent, H.~Dereli,
  V.~Dergachev, R.T. DeRosa, R.~De~Rosa, R.~DeSalvo, S.~Dhurandhar, M.C.
  D{\'{i}}az, L.~Di~Fiore, M.~Di~Giovanni, A.~Di~Lieto, S.~Di~Pace,
  I.~Di~Palma, A.~Di~Virgilio, G.~Dojcinoski, V.~Dolique, F.~Donovan, K.L.
  Dooley, S.~Doravari, R.~Douglas, T.P. Downes, M.~Drago, R.W.P. Drever, J.C.
  Driggers, Z.~Du, M.~Ducrot, S.E. Dwyer, T.B. Edo, M.C. Edwards, A.~Effler,
  H.B. Eggenstein, P.~Ehrens, J.~Eichholz, S.S. Eikenberry, W.~Engels, R.C.
  Essick, T.~Etzel, M.~Evans, T.M. Evans, R.~Everett, M.~Factourovich,
  V.~Fafone, H.~Fair, S.~Fairhurst, X.~Fan, Q.~Fang, S.~Farinon, B.~Farr, W.M.
  Farr, M.~Favata, M.~Fays, H.~Fehrmann, M.M. Fejer, D.~Feldbaum, I.~Ferrante,
  E.C. Ferreira, F.~Ferrini, F.~Fidecaro, L.S. Finn, I.~Fiori, D.~Fiorucci,
  R.P. Fisher, R.~Flaminio, M.~Fletcher, H.~Fong, J.D. Fournier, S.~Franco,
  S.~Frasca, F.~Frasconi, M.~Frede, Z.~Frei, A.~Freise, R.~Frey, V.~Frey, T.T.
  Fricke, P.~Fritschel, V.V. Frolov, P.~Fulda, M.~Fyffe, H.A.G. Gabbard, J.R.
  Gair, L.~Gammaitoni, S.G. Gaonkar, F.~Garufi, A.~Gatto, G.~Gaur, N.~Gehrels,
  G.~Gemme, B.~Gendre, E.~Genin, A.~Gennai, J.~George, L.~Gergely, V.~Germain,
  A.~Ghosh, A.~Ghosh, S.~Ghosh, J.A. Giaime, K.D. Giardina, A.~Giazotto,
  K.~Gill, A.~Glaefke, J.R. Gleason, E.~Goetz, R.~Goetz, L.~Gondan,
  G.~Gonz{\'{a}}lez, J.M.G. Castro, A.~Gopakumar, N.A. Gordon, M.L. Gorodetsky,
  S.E. Gossan, M.~Gosselin, R.~Gouaty, C.~Graef, P.B. Graff, M.~Granata,
  A.~Grant, S.~Gras, C.~Gray, G.~Greco, A.C. Green, R.J.S. Greenhalgh,
  P.~Groot, H.~Grote, S.~Grunewald, G.M. Guidi, X.~Guo, A.~Gupta, M.K. Gupta,
  K.E. Gushwa, E.K. Gustafson, R.~Gustafson, J.J. Hacker, B.R. Hall, E.D. Hall,
  G.~Hammond, M.~Haney, M.M. Hanke, J.~Hanks, C.~Hanna, M.D. Hannam, J.~Hanson,
  T.~Hardwick, J.~Harms, G.M. Harry, I.W. Harry, M.J. Hart, M.T. Hartman, C.J.
  Haster, K.~Haughian, J.~Healy, J.~Heefner, A.~Heidmann, M.C. Heintze,
  G.~Heinzel, H.~Heitmann, P.~Hello, G.~Hemming, M.~Hendry, I.S. Heng,
  J.~Hennig, A.W. Heptonstall, M.~Heurs, S.~Hild, D.~Hoak, K.A. Hodge,
  D.~Hofman, S.E. Hollitt, K.~Holt, D.E. Holz, P.~Hopkins, D.J. Hosken,
  J.~Hough, E.A. Houston, E.J. Howell, Y.M. Hu, S.~Huang, E.A. Huerta, D.~Huet,
  B.~Hughey, S.~Husa, S.H. Huttner, T.~Huynh-Dinh, A.~Idrisy, N.~Indik, D.R.
  Ingram, R.~Inta, H.N. Isa, J.M. Isac, M.~Isi, G.~Islas, T.~Isogai, B.R. Iyer,
  K.~Izumi, M.B. Jacobson, T.~Jacqmin, H.~Jang, K.~Jani, P.~Jaranowski,
  S.~Jawahar, F.~Jim{\'{e}}nez-Forteza, W.W. Johnson, N.K. Johnson-McDaniel,
  D.I. Jones, R.~Jones, R.J.G. Jonker, L.~Ju, K.~Haris, C.V. Kalaghatgi,
  V.~Kalogera, S.~Kandhasamy, G.~Kang, J.B. Kanner, S.~Karki, M.~Kasprzack,
  E.~Katsavounidis, W.~Katzman, S.~Kaufer, T.~Kaur, K.~Kawabe, F.~Kawazoe,
  F.~K{\'{e}}f{\'{e}}lian, M.S. Kehl, D.~Keitel, D.B. Kelley, W.~Kells,
  R.~Kennedy, D.G. Keppel, J.S. Key, A.~Khalaidovski, F.Y. Khalili, I.~Khan,
  S.~Khan, Z.~Khan, E.A. Khazanov, N.~Kijbunchoo, C.~Kim, J.~Kim, K.~Kim, N.G.
  Kim, N.~Kim, Y.M. Kim, E.J. King, P.J. King, D.L. Kinzel, J.S. Kissel,
  L.~Kleybolte, S.~Klimenko, S.M. Koehlenbeck, K.~Kokeyama, S.~Koley,
  V.~Kondrashov, A.~Kontos, S.~Koranda, M.~Korobko, W.Z. Korth, I.~Kowalska,
  D.B. Kozak, V.~Kringel, B.~Krishnan, A.~Kr{\'{o}}lak, C.~Krueger, G.~Kuehn,
  P.~Kumar, R.~Kumar, L.~Kuo, A.~Kutynia, P.~Kwee, B.D. Lackey, M.~Landry,
  J.~Lange, B.~Lantz, P.D. Lasky, A.~Lazzarini, C.~Lazzaro, P.~Leaci,
  S.~Leavey, E.O. Lebigot, C.H. Lee, H.K. Lee, H.M. Lee, K.~Lee, A.~Lenon,
  M.~Leonardi, J.R. Leong, N.~Leroy, N.~Letendre, Y.~Levin, B.M. Levine, T.G.F.
  Li, A.~Libson, T.B. Littenberg, N.A. Lockerbie, J.~Logue, A.L. Lombardi, L.T.
  London, J.E. Lord, M.~Lorenzini, V.~Loriette, M.~Lormand, G.~Losurdo, J.D.
  Lough, C.O. Lousto, G.~Lovelace, H.~L{\"{u}}ck, A.P. Lundgren, J.~Luo,
  R.~Lynch, Y.~Ma, T.~MacDonald, B.~Machenschalk, M.~MacInnis, D.M. Macleod,
  F.~Maga{\~{n}}a-Sandoval, R.M. Magee, M.~Mageswaran, E.~Majorana,
  I.~Maksimovic, V.~Malvezzi, N.~Man, I.~Mandel, V.~Mandic, V.~Mangano, G.L.
  Mansell, M.~Manske, M.~Mantovani, F.~Marchesoni, F.~Marion, S.~M{\'{a}}rka,
  Z.~M{\'{a}}rka, A.S. Markosyan, E.~Maros, F.~Martelli, L.~Martellini, I.W.
  Martin, R.M. Martin, D.V. Martynov, J.N. Marx, K.~Mason, A.~Masserot, T.J.
  Massinger, M.~Masso-Reid, F.~Matichard, L.~Matone, N.~Mavalvala, N.~Mazumder,
  G.~Mazzolo, R.~McCarthy, D.E. McClelland, S.~McCormick, S.C. McGuire,
  G.~McIntyre, J.~McIver, D.J. McManus, S.T. McWilliams, D.~Meacher, G.D.
  Meadors, J.~Meidam, A.~Melatos, G.~Mendell, D.~Mendoza-Gandara, R.A. Mercer,
  E.~Merilh, M.~Merzougui, S.~Meshkov, C.~Messenger, C.~Messick, P.M. Meyers,
  F.~Mezzani, H.~Miao, C.~Michel, H.~Middleton, E.E. Mikhailov, L.~Milano,
  J.~Miller, M.~Millhouse, Y.~Minenkov, J.~Ming, S.~Mirshekari, C.~Mishra,
  S.~Mitra, V.P. Mitrofanov, G.~Mitselmakher, R.~Mittleman, A.~Moggi, M.~Mohan,
  S.R.P. Mohapatra, M.~Montani, B.C. Moore, C.J. Moore, D.~Moraru, G.~Moreno,
  S.R. Morriss, K.~Mossavi, B.~Mours, C.M. Mow-Lowry, C.L. Mueller, G.~Mueller,
  A.W. Muir, A.~Mukherjee, D.~Mukherjee, S.~Mukherjee, N.~Mukund, A.~Mullavey,
  J.~Munch, D.J. Murphy, P.G. Murray, A.~Mytidis, I.~Nardecchia,
  L.~Naticchioni, R.K. Nayak, V.~Necula, K.~Nedkova, G.~Nelemans, M.~Neri,
  A.~Neunzert, G.~Newton, T.T. Nguyen, A.B. Nielsen, S.~Nissanke, A.~Nitz,
  F.~Nocera, D.~Nolting, M.E.N. Normandin, L.K. Nuttall, J.~Oberling,
  E.~Ochsner, J.~O’Dell, E.~Oelker, G.H. Ogin, J.J. Oh, S.H. Oh, F.~Ohme,
  M.~Oliver, P.~Oppermann, R.J. Oram, B.~O’Reilly, R.~O’Shaughnessy, C.D.
  Ott, D.J. Ottaway, R.S. Ottens, H.~Overmier, B.J. Owen, A.~Pai, S.A. Pai,
  J.R. Palamos, O.~Palashov, C.~Palomba, A.~Pal-Singh, H.~Pan, Y.~Pan,
  C.~Pankow, F.~Pannarale, B.C. Pant, F.~Paoletti, A.~Paoli, M.A. Papa, H.R.
  Paris, W.~Parker, D.~Pascucci, A.~Pasqualetti, R.~Passaquieti, D.~Passuello,
  B.~Patricelli, Z.~Patrick, B.L. Pearlstone, M.~Pedraza, R.~Pedurand,
  L.~Pekowsky, A.~Pele, S.~Penn, A.~Perreca, H.P. Pfeiffer, M.~Phelps,
  O.~Piccinni, M.~Pichot, M.~Pickenpack, F.~Piergiovanni, V.~Pierro,
  G.~Pillant, L.~Pinard, I.M. Pinto, M.~Pitkin, J.H. Poeld, R.~Poggiani,
  P.~Popolizio, A.~Post, J.~Powell, J.~Prasad, V.~Predoi, S.S. Premachandra,
  T.~Prestegard, L.R. Price, M.~Prijatelj, M.~Principe, S.~Privitera, R.~Prix,
  G.A. Prodi, L.~Prokhorov, O.~Puncken, M.~Punturo, P.~Puppo, M.~P{\"{u}}rrer,
  H.~Qi, J.~Qin, V.~Quetschke, E.A. Quintero, R.~Quitzow-James, F.J. Raab, D.S.
  Rabeling, H.~Radkins, P.~Raffai, S.~Raja, M.~Rakhmanov, C.R. Ramet,
  P.~Rapagnani, V.~Raymond, M.~Razzano, V.~Re, J.~Read, C.M. Reed, T.~Regimbau,
  L.~Rei, S.~Reid, D.H. Reitze, H.~Rew, S.D. Reyes, F.~Ricci, K.~Riles, N.A.
  Robertson, R.~Robie, F.~Robinet, A.~Rocchi, L.~Rolland, J.G. Rollins, V.J.
  Roma, J.D. Romano, R.~Romano, G.~Romanov, J.H. Romie, D.~Rosi{\'{n}}ska,
  S.~Rowan, A.~R{\"{u}}diger, P.~Ruggi, K.~Ryan, S.~Sachdev, T.~Sadecki,
  L.~Sadeghian, L.~Salconi, M.~Saleem, F.~Salemi, A.~Samajdar, L.~Sammut, L.M.
  Sampson, E.J. Sanchez, V.~Sandberg, B.~Sandeen, G.H. Sanders, J.R. Sanders,
  B.~Sassolas, B.S. Sathyaprakash, P.R. Saulson, O.~Sauter, R.L. Savage,
  A.~Sawadsky, P.~Schale, R.~Schilling, J.~Schmidt, P.~Schmidt, R.~Schnabel,
  R.M.S. Schofield, A.~Sch{\"{o}}nbeck, E.~Schreiber, D.~Schuette, B.F. Schutz,
  J.~Scott, S.M. Scott, D.~Sellers, A.S. Sengupta, D.~Sentenac, V.~Sequino,
  A.~Sergeev, G.~Serna, Y.~Setyawati, A.~Sevigny, D.A. Shaddock, T.~Shaffer,
  S.~Shah, M.S. Shahriar, M.~Shaltev, Z.~Shao, B.~Shapiro, P.~Shawhan,
  A.~Sheperd, D.H. Shoemaker, D.M. Shoemaker, K.~Siellez, X.~Siemens, D.~Sigg,
  A.D. Silva, D.~Simakov, A.~Singer, L.P. Singer, A.~Singh, R.~Singh,
  A.~Singhal, A.M. Sintes, B.J.J. Slagmolen, J.R. Smith, M.R. Smith, N.D.
  Smith, R.J.E. Smith, E.J. Son, B.~Sorazu, F.~Sorrentino, T.~Souradeep, A.K.
  Srivastava, A.~Staley, M.~Steinke, J.~Steinlechner, S.~Steinlechner,
  D.~Steinmeyer, B.C. Stephens, S.P. Stevenson, R.~Stone, K.A. Strain,
  N.~Straniero, G.~Stratta, N.A. Strauss, S.~Strigin, R.~Sturani, A.L. Stuver,
  T.Z. Summerscales, L.~Sun, P.J. Sutton, B.L. Swinkels, M.J.
  Szczepa{\'{n}}czyk, M.~Tacca, D.~Talukder, D.B. Tanner, M.~T{\'{a}}pai, S.P.
  Tarabrin, A.~Taracchini, R.~Taylor, T.~Theeg, M.P. Thirugnanasambandam, E.G.
  Thomas, M.~Thomas, P.~Thomas, K.A. Thorne, K.S. Thorne, E.~Thrane, S.~Tiwari,
  V.~Tiwari, K.V. Tokmakov, C.~Tomlinson, M.~Tonelli, C.V. Torres, C.I. Torrie,
  D.~T{\"{o}}yr{\"{a}}, F.~Travasso, G.~Traylor, D.~Trifir{\`{o}}, M.C.
  Tringali, L.~Trozzo, M.~Tse, M.~Turconi, D.~Tuyenbayev, D.~Ugolini, C.S.
  Unnikrishnan, A.L. Urban, S.A. Usman, H.~Vahlbruch, G.~Vajente, G.~Valdes,
  M.~Vallisneri, N.~van Bakel, M.~van Beuzekom, J.F.J. van~den Brand, C.~Van
  Den~Broeck, D.C. Vander-Hyde, L.~van~der Schaaf, J.V. van Heijningen, A.A.
  van Veggel, M.~Vardaro, S.~Vass, M.~Vas{\'{u}}th, R.~Vaulin, A.~Vecchio,
  G.~Vedovato, J.~Veitch, P.J. Veitch, K.~Venkateswara, D.~Verkindt,
  F.~Vetrano, A.~Vicer{\'{e}}, S.~Vinciguerra, D.J. Vine, J.Y. Vinet,
  S.~Vitale, T.~Vo, H.~Vocca, C.~Vorvick, D.~Voss, W.D. Vousden, S.P.
  Vyatchanin, A.R. Wade, L.E. Wade, M.~Wade, S.J. Waldman, M.~Walker,
  L.~Wallace, S.~Walsh, G.~Wang, H.~Wang, M.~Wang, X.~Wang, Y.~Wang, H.~Ward,
  R.L. Ward, J.~Warner, M.~Was, B.~Weaver, L.W. Wei, M.~Weinert, A.J.
  Weinstein, R.~Weiss, T.~Welborn, L.~Wen, P.~We{\ss}els, T.~Westphal,
  K.~Wette, J.T. Whelan, S.E. Whitcomb, D.J. White, B.F. Whiting, K.~Wiesner,
  C.~Wilkinson, P.A. Willems, L.~Williams, R.D. Williams, A.R. Williamson, J.L.
  Willis, B.~Willke, M.H. Wimmer, L.~Winkelmann, W.~Winkler, C.C. Wipf, A.G.
  Wiseman, H.~Wittel, G.~Woan, J.~Worden, J.L. Wright, G.~Wu, J.~Yablon,
  I.~Yakushin, W.~Yam, H.~Yamamoto, C.C. Yancey, M.J. Yap, H.~Yu, M.~Yvert,
  A.~Zadro{\.{z}}ny, L.~Zangrando, M.~Zanolin, J.P. Zendri, M.~Zevin, F.~Zhang,
  L.~Zhang, M.~Zhang, Y.~Zhang, C.~Zhao, M.~Zhou, Z.~Zhou, X.J. Zhu, M.E.
  Zucker, S.E. Zuraw, J.~Zweizig, Physical Review Letters \textbf{116}(6),
  061102 (2016).
\newblock \doi{10.1103/PhysRevLett.116.061102}

\bibitem{Will2014}
C.M. Will, Living Reviews in Relativity \textbf{17} (2014).
\newblock \doi{10.12942/lrr-2014-4}.
\newblock \urlprefix\url{http://www.livingreviews.org/lrr-2014-4}

\bibitem{Demorest2010}
P.B. Demorest, T.~Pennucci, S.M. Ransom, M.S.E. Roberts, J.W.T. Hessels, Nature
  \textbf{467}(7319), 1081 (2010).
\newblock \doi{10.1038/nature09466}.
\newblock
  \urlprefix\url{https://www.nature.com/nature/journal/v467/n7319/full/nature09466.html
  https://www.nature.com/nature/journal/v467/n7319/pdf/nature09466.pdf
  http://www.nature.com/doifinder/10.1038/nature09466
  http://arxiv.org/abs/1010.5788 http://dx.doi.org/10.1038/nature}

\bibitem{Berti2005}
E.~Berti, A.~Buonanno, C.M. Will, Classical and Quantum Gravity
  \textbf{22}(18), S943 (2005).
\newblock \doi{10.1088/0264-9381/22/18/S08}.
\newblock \urlprefix\url{http://stacks.iop.org/0264-9381/22/i=18/a=S08
  http://iopscience.iop.org/article/10.1088/0264-9381/22/18/S08/pdf
  http://arxiv.org/abs/gr-qc/0504017
  http://dx.doi.org/10.1088/0264-9381/22/18/S08}

\bibitem{Moraes2014a}
P.H.R.S. Moraes, O.D. Miranda, Astrophysics and Space Science \textbf{354}(2),
  2121 (2014).
\newblock \doi{10.1007/s10509-014-2121-6}.
\newblock
  \urlprefix\url{https://link.springer.com/article/10.1007/s10509-014-2121-6
  https://link.springer.com/content/pdf/10.1007%2Fs10509-014-2121-6.pdf
  http://link.springer.com/10.1007/s10509-014-2121-6}

\bibitem{Das2015}
U.~Das, B.~Mukhopadhyay, International Journal of Modern Physics D
  \textbf{24}(12), 1544026 (2015).
\newblock \doi{10.1142/S0218271815440265}.
\newblock
  \urlprefix\url{http://www-worldscientific-com.ez63.periodicos.capes.gov.br/doi/abs/10.1142/S0218271815440265
  http://www-worldscientific-com.ez63.periodicos.capes.gov.br/doi/pdf/10.1142/S0218271815440265}

\bibitem{Jain2016}
R.K. Jain, C.~Kouvaris, N.G. Nielsen, Physical Review Letters \textbf{116}(15),
  1 (2016).
\newblock \doi{10.1103/PhysRevLett.116.151103}

\bibitem{Harko2011}
T.~Harko, F.S.N. Lobo, S.~Nojiri, S.D. Odintsov, Physical Review D
  \textbf{84}(2), 024020 (2011).
\newblock \doi{10.1103/PhysRevD.84.024020}.
\newblock
  \urlprefix\url{http://link.aps.org/doi/10.1103/PhysRevD.84.024020%0Ahttps://journals.aps.org.ez63.periodicos.capes.gov.br/prd/abstract/10.1103/PhysRevD.84.024020%0Ahttps://journals.aps.org.ez63.periodicos.capes.gov.br/prd/pdf/10.1103/PhysRevD.84.024020
  https://link.aps.}

\bibitem{Tolman1939}
R.C. Tolman, Physical Review \textbf{55}(4), 364 (1939).
\newblock \doi{10.1103/PhysRev.55.364}.
\newblock \urlprefix\url{https://link.aps.org/doi/10.1103/PhysRev.55.364
  https://journals.aps.org/pr/abstract/10.1103/PhysRev.55.364
  https://journals.aps.org/pr/pdf/10.1103/PhysRev.55.364}

\bibitem{Oppenheimer1939}
J.R. Oppenheimer, G.M. Volkoff, Physical Review \textbf{55}(4), 374 (1939).
\newblock \doi{10.1103/PhysRev.55.374}

\bibitem{Zaregonbadi2016}
R.~Zaregonbadi, M.~Farhoudi, N.~Riazi, Physical Review D \textbf{94}(8), 084052
  (2016).
\newblock \doi{10.1103/PhysRevD.94.084052}.
\newblock \urlprefix\url{https://link.aps.org/doi/10.1103/PhysRevD.94.084052}

\bibitem{Shabani2014}
H.~Shabani, M.~Farhoudi, Physical Review D \textbf{90}(4), 044031 (2014).
\newblock \doi{10.1103/PhysRevD.90.044031}.
\newblock \urlprefix\url{https://link.aps.org/doi/10.1103/PhysRevD.90.044031
  https://journals.aps.org/prd/abstract/10.1103/PhysRevD.90.044031
  https://journals.aps.org/prd/pdf/10.1103/PhysRevD.90.044031
  http://arxiv.org/abs/1407.6187 http://dx.doi.org/10.1103/PhysRevD.90.044031}

\bibitem{Moraes2016d}
P.H.R.S. Moraes, J.R.L. Santos, The European Physical Journal C \textbf{76}(2),
  60 (2016).
\newblock \doi{10.1140/epjc/s10052-016-3912-4}.
\newblock
  \urlprefix\url{https://link.springer.com/article/10.1140/epjc/s10052-016-3912-4
  https://link.springer.com/content/pdf/10.1140%2Fepjc%2Fs10052-016-3912-4.pdf
  http://arxiv.org/abs/1601.02811
  http://dx.doi.org/10.1140/epjc/s10052-016-3912-4
  http://link.springer.com/10.1140}

\bibitem{Sharif2012}
M.~Sharif, M.~Zubair, Journal of Cosmology and Astroparticle Physics
  \textbf{2012}(03), 028 (2012).
\newblock \doi{10.1088/1475-7516/2012/03/028}.
\newblock \urlprefix\url{http://stacks.iop.org/1475-7516/2012/i=03/a=028
  http://iopscience.iop.org/article/10.1088/1475-7516/2012/03/028/pdf
  http://stacks.iop.org/1475-7516/2012/i=03/a=028?key=crossref.3d96d96a4566a0dc35c20081b4b8d51f}

\bibitem{Nojiri2011}
S.~Nojiri, S.D. Odintsov, Physics Reports \textbf{505}(2–4), 59 (2011).
\newblock \doi{10.1016/j.physrep.2011.04.001}.
\newblock
  \urlprefix\url{http://www.sciencedirect.com/science/article/pii/S0370157311001335
  https://www.sciencedirect.com/science/article/pii/S0370157311001335}

\bibitem{Alvarenga2013a}
F.G. Alvarenga, A.~de~la Cruz-Dombriz, M.J.S. Houndjo, M.E. Rodrigues,
  D.~S{\'{a}}ez-G{\'{o}}mez, Physical Review D \textbf{87}(10), 103526 (2013).
\newblock \doi{10.1103/PhysRevD.87.103526}.
\newblock \urlprefix\url{https://link.aps.org/doi/10.1103/PhysRevD.87.103526
  https://journals.aps.org/prd/abstract/10.1103/PhysRevD.87.103526
  https://journals.aps.org/prd/pdf/10.1103/PhysRevD.87.103526}

\bibitem{BarrientosO.2014}
J.~Barrientos~O., G.F. Rubilar, Physical Review D \textbf{90}(2), 028501
  (2014).
\newblock \doi{10.1103/PhysRevD.90.028501}.
\newblock \urlprefix\url{https://link.aps.org/doi/10.1103/PhysRevD.90.028501
  https://journals.aps.org/prd/abstract/10.1103/PhysRevD.90.028501
  https://journals.aps.org/prd/pdf/10.1103/PhysRevD.90.028501}

\bibitem{Moraes2016}
P.H.R.S. Moraes, J.D.V. Arba{\~{n}}il, M.~Malheiro, Journal of Cosmology and
  Astroparticle Physics \textbf{2016}(06), 005 (2016).
\newblock \doi{10.1088/1475-7516/2016/06/005}.
\newblock \urlprefix\url{http://stacks.iop.org/1475-7516/2016/i=06/a=005}

\bibitem{Alvarenga2013}
F.G. Alvarenga, M.J.S. Houndjo, A.V. Monwanou, J.B.C. Orou, Journal of Modern
  Physics \textbf{04}(01), 130 (2013).
\newblock \doi{10.4236/jmp.2013.41019}.
\newblock
  \urlprefix\url{http://www.scirp.org/journal/PaperInformation.aspx?PaperID=27253&#abstract
  http://www.scirp.org/journal/PaperDownload.aspx?paperID=27253
  http://www.scirp.org/journal/PaperInformation.aspx?paperID=27253}

\bibitem{Moraes2014}
P.H.R.S. Moraes, Astrophysics and Space Science \textbf{352}(1), 273 (2014).
\newblock \doi{10.1007/s10509-014-1895-x}.
\newblock
  \urlprefix\url{https://link.springer.com/article/10.1007/s10509-014-1895-x
  https://link.springer.com/content/pdf/10.1007%2Fs10509-014-1895-x.pdf}

\bibitem{Moraes2015}
P.H.R.S. Moraes, The European Physical Journal C \textbf{75}(4), 168 (2015).
\newblock \doi{10.1140/epjc/s10052-015-3393-x}.
\newblock
  \urlprefix\url{https://link.springer.com/article/10.1140/epjc/s10052-015-3393-x
  https://link.springer.com/content/pdf/10.1140%2Fepjc%2Fs10052-015-3393-x.pdf}

\bibitem{Shamir2015}
M.F. Shamir, The European Physical Journal C \textbf{75}(8), 354 (2015).
\newblock \doi{10.1140/epjc/s10052-015-3582-7}.
\newblock
  \urlprefix\url{https://link.springer.com/article/10.1140/epjc/s10052-015-3582-7
  https://link.springer.com/content/pdf/10.1140%2Fepjc%2Fs10052-015-3582-7.pdf}

\bibitem{Moraes2016e}
P.H.R.S. Moraes, International Journal of Theoretical Physics \textbf{55}(3),
  1307 (2016).
\newblock \doi{10.1007/s10773-015-2771-3}.
\newblock
  \urlprefix\url{https://link.springer.com/article/10.1007/s10773-015-2771-3
  https://link.springer.com/content/pdf/10.1007%2Fs10773-015-2771-3.pdf}

\bibitem{Alves2016}
M.E.S. Alves, P.H.R.S. Moraes, J.C.N. de~Araujo, M.~Malheiro, Physical Review D
  \textbf{94}(2), 024032 (2016).
\newblock \doi{10.1103/PhysRevD.94.024032}.
\newblock \urlprefix\url{https://link.aps.org/doi/10.1103/PhysRevD.94.024032
  https://journals.aps.org/prd/abstract/10.1103/PhysRevD.94.024032}

\bibitem{Moraes2016a}
P.H.R.S. Moraes, R.A.C. Correa, R.V. Lobato, arXiv:1701.01028 pp. 1--14 (2016).
\newblock \urlprefix\url{http://arxiv.org/abs/1701.01028}

\bibitem{Chandrasekhar1935}
S.~Chandrasekhar, Monthly Notices of the Royal Astronomical Society
  \textbf{95}(3), 207 (1935).
\newblock \doi{10.1093/mnras/95.3.207}.
\newblock
  \urlprefix\url{https://academic.oup.com/mnras/article/95/3/207/991812/The-Highly-Collapsed-Configurations-of-a-Stellar
  https://academic.oup.com/mnras/article-lookup/doi/10.1093/mnras/95.3.207
  https://academic.oup.com/mnras/article-pdf/95/3/207/6181203/mnras95-0207.pdf}

\bibitem{Vennes1997}
S.~Vennes, P.A. Thejll, R.G. Galvan, J.~Dupuis, The Astrophysical Journal
  \textbf{480}(1986), 714 (1997).
\newblock \doi{10.1086/303981}

\bibitem{Haensel2004}
M.~Nale{\.{z}}yty, J.~Madej, Astronomy {\&} Astrophysics \textbf{420}(2), 507
  (2004).
\newblock \doi{10.1051/0004-6361:20040123}.
\newblock \urlprefix\url{http://arxiv.org/abs/astro-ph/0603713
  http://dx.doi.org/10.1051/0004-6361:20064956
  http://www.aanda.org/10.1051/0004-6361:20064956
  http://www.aanda.org/10.1051/0004-6361:20040123}

\bibitem{Wehbring2016}
J.W.E. Wehbring, D.~Bateman, S.~Hauberg, Rik.
\newblock {{\{}GNU Octave{\}} version 4.2.0 manual: a high-level interactive
  language for numerical computations} (2016).
\newblock \doi{http://www.gnu.org/software/octave/doc/interpreter}.
\newblock \urlprefix\url{http://www.gnu.org/software/octave/doc/interpreter}

\bibitem{SageMath}
{The Developers Sage}.
\newblock {{\{}S{\}}ageMath, the {\{}S{\}}age {\{}M{\}}athematics
  {\{}S{\}}oftware {\{}S{\}}ystem ({\{}V{\}}ersion 7.6.0)} (2017).
\newblock \doi{http://www.sagemath.org}.
\newblock \urlprefix\url{http://www.sagemath.org}

\bibitem{Droettboom2017}
M.~Droettboom, T.A. Caswell, J.~Hunter, E.~Firing, J.H. Nielsen, B.~Root,
  P.~Elson, D.~Dale, J.J. Lee, N.~Varoquaux, J.K. Sepp��nen, D.~McDougall,
  R.~May, A.~Straw, E.S.D. Andrade, A.~Lee, T.S. Yu, E.~Ma, C.~Gohlke,
  S.~Silvester, C.~Moad, P.~Hobson, J.~Schulz, P.~W��rtz, F.~Ariza,
  {Cimarron}, T.~Hisch, N.~Kniazev, A.F. Vincent, I.~Thomas,   (2017).
\newblock \doi{10.5281/ZENODO.248351}.
\newblock \urlprefix\url{https://zenodo.org/record/248351}

\bibitem{Python}
{Python Software Foundation}.
\newblock {Python Language Reference, version 3.6.0} (2017).
\newblock \doi{https://www.python.org/}.
\newblock \urlprefix\url{https://www.python.org/}

\bibitem{Maxima}
{Maxima}.
\newblock {Maxima, a Computer Algebra System. Version 5.39.0} (2016).
\newblock \doi{http://maxima.sourceforge.net/}.
\newblock \urlprefix\url{http://maxima.sourceforge.net/}

\end{thebibliography}


\begin{thebibliography}{}

\bibitem{Woosley2015}
S.E. Woosley, A. Heger, Astrophys. J. {\bf 810}, 34 (2015).

\bibitem{Chandrasekhar1931}
S. Chandrasekhar, Astrophys. J. {\bf 74}, 81 (1931).

\bibitem{Weinberg2008}
S. Weinberg, {\it Cosmology}. (OUP, Oxford, 2008).

\bibitem{Riess1998}
A.G. Riess {\it et al.}, Astron. J {\bf 116}(3), 1009-1038 (1998).

\bibitem{Perlmutter1999}
S. Perlmutter {\it et al.}, Astrophys. J. {\bf 517}(2), 565 (1999).

\bibitem{AndrewHowell2006}
D.A. Howell {\it et al.}, Nature, {\bf 443}, 308-311 (2006).

\bibitem{Scalzo2010}
R.A. Scalzo {\it et al.}, Astrophys. J. {\bf 713}(2), 1073 (2010).

\bibitem{Hicken2007}
M. Hicken {\it et al.}, Astrophys. J. Lett. {\bf 669}(1), L17 (2007).

\bibitem{Yamanaka2009}
M. Yamanaka {\it et al.}, Astrophys. J. Lett. {\bf 707}(2), L118 (2009).

\bibitem{Taubenberger2011}
S. Taubenberger {\it et al.}, Mon. Not. R. Astron. Soc. {\bf 412}(4), 2735-2762 (2011).

\bibitem{Silverman2011}
J.M. Silverman {\it et al.}, Mon. Not. R. Astron. Soc. {\bf 410}(1), 585-611 (2011).

\bibitem{Boshkayev2013a}
K. Boshkayev {\it et al.}, Astrophys. J. {\bf 762}(2), 117 (2012).

\bibitem{Bera2016}
P. Bera, D. Bhattacharya, Mon. Not. R. Astron. Soc. {\bf 456}(3), 3375-3385 (2016).

\bibitem{Jaziel2014}
J.G. Coelho {\it et al.}, Astrophys. J. {\bf 794}(10), 86 (2016).

\bibitem{Franzon2015}
B. Franzon, S. Schramm, Phys. Rev. D {\bf 92}(10), 083006 (2015).

\bibitem{Otoniel2016}
E. Otoniel {\it et al.} \url{arXiv:1609.05994}

\bibitem{Deu2014}
W. De-Hua, L. He-Lei, Z. Xiang-Dong, Chin. Phys. B {\bf 23}(8), 089501 (2014).

\bibitem{Ji2013}
S. Ji {\it et al.}, Astrophys. J. {\bf 773}(2), 136 (2013).

\bibitem{ChandraTooper}
S. Chandrasekhar, R.F. Tooper, Astrophys. J. {\bf 139}, 1396 (1964).

\bibitem{Carvalho2015}
G. Carvalho, R. Marinho, M. Malheiro, J. Phys. Conf. Ser., {\bf 630}(1), 012058 (2015).

\bibitem{Carvalho2015a}
G. Carvalho, R. Marinho, M. Malheiro, in {\it THE SECOND ICRANet C\'ESAR LATTES MEETING: Supernovae, Neutron Stars and Black Holes}, vol. 1693, p. 030004. AIP Publishing (2015).

\bibitem{Carvalho2017}
G.A. Carvalho, R.M. Marinho Jr, M.Malheiro. \url{arXiv:1709.01635}

\bibitem{LIGOScientificCollaborationandVirgoCollaboration2016}
B.P. Abbott {\it et al.}, Phys. Rev. Lett. {\bf 116}(6), 061102 (2016).

\bibitem{Will2014}
C.M. Will, Living Rev. Relat. {\bf 17}(1), 4 (2014).

\bibitem{Demorest2010}
P.B. Demorest {\it et al.}, Nature {\bf 467}(7319), 1081-1083 (2010).

\bibitem{Berti2005}
E. Berti, A. Buonanno, C.M. Will, Class. Quantum Gravity {\bf 22}(18), S943 (2005).

\bibitem{Moraes2014a}
P.H.R.S. Moraes, O.D. Miranda, Astrophys. Space Sci. {\bf 354}(2), 2121 (2014).

\bibitem{Das2015}
U. Das, B. Mukhopadhyay, Int. J. Mod. Phys. D {\bf 24}(12), 1544026 (2015).

\bibitem{Das2015a}
U. Das, B. Mukhopadhyay, J. Cosmol. Astropart. Phys. {\bf 2015}(05), 045 (2015).

\bibitem{Jain2016}
R.K. Jain, C. Kouvaris, N.G. Nielsen, Phys. Rev. Lett. {\bf 116}(15), 151103 (2016).

\bibitem{Harko2011}
T. Harko {\it et al.}, Phys. Rev. D {\bf 84}(2), 024020 (2011).
  
\bibitem{Tolman1939}
R.C. Tolman, Phys. Rev. {\bf 55}(4), 364-373 (1939).

\bibitem{Oppenheimer1939}
J.R. Oppenheimer, G.M. Volkoff, Phys. Rev. {\bf 55}(4), 374-381 (1939).

\bibitem{Zaregonbadi2016}
R. Zaregonbadi, M. Farhoudi, N. Riazi, Phys. Rev. D {\bf 94}(8), 084052 (2016).

\bibitem{Shabani2014}
H. Shabani, M. Farhoudi, Phys. Rev. D {\bf 90}(4), 044031 (2014).

\bibitem{Moraes2016d}
P.H.R.S. Moraes, J.R.L. Santos, Eur. Phys. J. C {\bf 76}(2), 60 (2016).

\bibitem{ms/2017}
P.H.R.S. Moraes, P.K. Sahoo, Eur. Phys. J. C {\bf 77}(7), 480 (2017).

\bibitem{Sharif2012}
M. Sharif, M. Zubair, J. Cosmol. Astropart. Phys. {\bf 2012}(03), 028 (2012).

\bibitem{Moraes2016}
P.H.R.S. Moraes, J.D.V. Arbanil, M. Malheiro, J. Cosmol. Astropart. Phys. {\bf 2016}(06), 005 (2016).

\bibitem{das/2016}
A. Das {\it et al.}, Eur. Phys. J. C {\bf 76}(12), 654 (2016).

\bibitem{sharif/2014}
M. Sharif, Z. Yousaf, Astrophys. Space Sci. {\bf 354}(2), 2113 (2014).

\bibitem{noureen/2015}
I. Noureen, M. Zubair, Astrophys. Space Sci. {\bf 356}(1), 103-110 (2015).

\bibitem{noureen/2015b}
I. Noureen, M. Zubair, Eur. Phys. J. C {\bf 75}(2), 62 (2015).

\bibitem{noureen/2015c}
I. Noureen {\it et al.}, Eur. Phys. J. C {\bf 75}(7), 323 (2015).

\bibitem{Zubair2015} 
M. Zubair, I. Noureen, Eur. Phys. J. C {\bf 75}(6), 265 (2015).  

\bibitem{Nojiri2011}
S. Nojiri, S.D. Odintsov, Phys. Rep. {\bf 505}(2), 59-144 (2011).

\bibitem{Alvarenga2013a}
F.G. Alvarenga {\it et al.}, Phys. Rev. D, {\bf 87}(10), 103526 (2013).

\bibitem{BarrientosO.2014}
J. Barrientos O., G.F. Rubilar, Phys. Rev. D, {\bf 90}(2), 028501 (2014).

\bibitem{Alvarenga2013}
F.G. Alvarenga {\it et al.}, Journal of Modern Physics, {\bf 04}(01), 130-139 (2013).

\bibitem{Moraes2014}
P.H.R.S. Moraes, Astrophys. Space Sci. {\bf 352}(1), 273-279 (2014).

\bibitem{Moraes2015}
P.H.R.S. Moraes, Eur. Phys. J. C {\bf 75}(4), 168 (2015).

\bibitem{Shamir2015}
M.F. Shamir, Eur. Phys. J. C {\bf 75}(8), 354 (2015).

\bibitem{Moraes2016a}
P.H.R.S. Moraes, R.A.C. Correa, R.V. Lobato, J. Cosmol. Astropart. Phys. {\bf 2017}(07), 029 (2017).

\bibitem{Chandrasekhar1935}
S. Chandrasekhar, Mon. Not. R. Astron. Soc. {\bf 95}, 207-225 (1935).

\bibitem{Vennes1997}
S. Vennes, P.A. Thejll, R.G. Galvan, J. Dupuis, Astrophys. J. {\bf 480}, 714-734 (1997).

\bibitem{Nalezyty2004}
M. Nalezyty, J. Madej, Astron. Astrophys. {\bf 420}, 507-513 (2004).

\bibitem{Manuel2003}
M. Malheiro, M. Fiolhais, A.R. Taurines, J. Phys. G {\bf 29}, 1045 (2003).

\bibitem{banerjee/2017}
S. Banerjee {\it et al.} \url{arXiv:1705.01048}

\bibitem{althaus/2011}
L.G. Althaus {\it et al.}, Astron. Astrophys. {\bf 527}, A72 (2011).

\bibitem{garcia-berro/2011}
E. Garc\'ia-Berro {\it et al.}, J. Cosmol. Astropart. Phys. {\bf 2011}(05), 021 (2011).

\bibitem{corsico/2013}
A.H. C\'orsico {\it et al.}, J. Cosmol. Astropart. Phys. {\bf 2013}(06), 032 (2013).

\bibitem{mikheev/2016}
S.A. Mikheev and V.P. Tsvetkov, Phys. Part. Nucl. Lett. {\bf 13}(4), 442-450 (2016).

\bibitem{hamada/1961}
T. Hamada and E.E. Salpeter, Astrophys. J. {\bf 134}, 683 (1961).

\bibitem{Chamel2013}
N. Chamel, A.F. Fantina, P.J. Davis, Phys. Rev. D {\bf 88}, 081301(R) (2013).

\bibitem{harko/2014}
T. Harko et al., Phys. Rev. D {\bf 89}, 124036 (2014).

\bibitem{harko/2014b}
T. Harko et al., JCAP {\bf 12}, 021 (2014).

\bibitem{pace/2017} M. Pace and J.L. Said, Eur. Phys. J. C {\bf 77}, 62 (2017). 

\bibitem{Wehbring2016}
J. W. Eaton {\it et al.}, GNU Octave version 4.2.0, (2016). \url{http://www.gnu.org/software/octave/doc/interpreter}

\bibitem{Droettboom2017}
M. Droettboom {\it et al.}, Matplotlib: V2.0.0 (2017). \url{https://doi.org/10.5281/zenodo.248351}

\bibitem{Python}
Python Software Foundation: Python Language Reference, version 3.6.0 (2017). \url{https://www.python.org/}

\bibitem{Maxima}
Maxima, a Computer Algebra System. Version 5.39.0 (2016). \url{http://maxima.sourceforge.net/}
\end{thebibliography}
\end{document}